\documentclass[12pt,twoside]{article}
\usepackage[latin9]{inputenc}
\usepackage{array}
\usepackage{amsmath}
\usepackage{graphicx}
\usepackage[unicode=true,pdfusetitle,
bookmarks=true,bookmarksnumbered=false,bookmarksopen=false,
 breaklinks=true,pdfborder={0 0 0},pdfborderstyle={},backref=false,colorlinks=true,allcolors=blue] 
 {hyperref}

\makeatletter

\providecommand{\tabularnewline}{\\}

\usepackage{correl}

\def\TheTitle{Dynamical mean field theory with quantum computing}
\def\TheHeading{Dynamical mean field theory with quantum computing}
\def\TheAuthor{Thomas Ayral}
\def\TheInstitute{Eviden Quantum Laboratory}
\def\TheAddress{Les Clayes-sous-Bois, France}
\def\TheChapter{5}           

\makeatother

\begin{document}
\MakeTitle 

\section{Introduction: why quantum computers for dynamical mean field theory?}

Quantum computers were originally proposed to solve the challenges
met by classical computers to tackle the quantum many-body problem
\cite{Feynman1982}. One famous representative of these problems is
the Fermi-Hubbard model \cite{Hubbard1963}
\begin{equation}
H=\sum_{ij,\sigma}t_{ij}c_{i\sigma}^{\dagger}c_{j\sigma}+U\sum_{i}n_{i\uparrow}n_{i\downarrow}-\mu\sum_{i}n_{i},\label{eq:hubbard_model}
\end{equation}
which describes the competition between a kinetic term with hopping
amplitudes $t_{ij}$ that favors delocalized states, and an interaction
term with interaction strength $U$ that favors localized states,
with extra complexity coming from the possibility to tune the average
density using a chemical potential $\mu$. Here, creation (resp. annihilation)
operators $c_{i\sigma}^{\dagger}$ (resp. $c_{i\sigma}$) create electrons
of spin $\sigma=\uparrow,\downarrow$ on lattice site $i$, and $n_{i\sigma}=c_{i\sigma}^{\dagger}c_{i\sigma}$,
$n_{i}=n_{i\uparrow}+n_{i\downarrow}$. Among other reasons, the purported
relation between this model and the physics of high-temperature cuprate
superconductors spurred early and prolonged interest into the phase
diagram of this model, with phases as diverse as Fermi liquids, Mott
insulators, superconductors or charge density insulators.

In particular, one central object to elucidate these phase diagrams
is the so-called spectral function $A(\boldsymbol{k},\omega)$, a
quantity that is accessible via e.g angle-resolved photoemission experiments:
its momentum $\boldsymbol{k}$ and energy $\omega$ dependence contains
distinctive features of the aforementioned phases. From a computational
point of view, the spectral function can be computed from the imaginary
part of the retarded Green's function $G^{\mathrm{R}}(\boldsymbol{k},\omega)$
($A(\boldsymbol{k},\omega)=-\frac{1}{\pi}\mathrm{Im}G^{\mathrm{R}}(\boldsymbol{k},\omega)$),
itself defined as the space and time Fourier transform of the following
sum of correlation functions\index{Green's function}:
\begin{equation}
G^{\mathrm{R}}(i,j;t)=\theta(t)\left(-i\langle c_{i}(t)c_{j}^{\dagger}\rangle-i\langle c_{j}^{\dagger}c_{i}(t)\rangle\right).\label{eq:retarded_gf}
\end{equation}
Here, the average denotes $\langle\cdots\rangle=\frac{1}{Z}\mathrm{Tr}\left(e^{-\beta H}\cdots\right)$,
with the partition function $Z=\mathrm{Tr}(e^{-\beta H})$ and the
inverse temperature $\beta=1/T$. Time-dependence is to be understood
in the Heisenberg picture, $c_{i}(t)=e^{iHt}c_{i}e^{-iHt}$. In the
zero-temperature limit, the average becomes $\langle\cdots\rangle=\langle\Psi_{0}|\cdots|\Psi_{0}\rangle$,
with $|\Psi_{0}\rangle$ the ground state of $H$. The retarded Green's
function describes how an electron created at site $j$ ($c_{j}^{\dagger}$)
in the system (described by a Gibbs state at finite temperature or
the ground state at zero temperature) propagates to site $i$ for
a time $t$, where it is annihilated (and likewise for a hole, described
by the second term). From this definition, one sees that computing
$A(\boldsymbol{k},\omega)$ implies the ability to describe the time
evolution of an electron or a hole in a quantum system prepared in
its Gibbs or ground state.

\subsection{The difficulties of classical methods... and of quantum computers}

Due to the many-body nature of the problem, traditional mean-field
theories fail to properly describe these states and their subsequent
evolution: for instance, the Hartree-Fock method does not capture
Mott insulators because they cannot be described with single Slater
determinants. In other words, Hubbard physics is generally not described
well by single-particle physics: correlations (essentially entanglement
beyond the trivial entanglement required by the Pauli principle) play
an important role. This warrants the use of more sophisticated classical
methods. These are either exact, but with a cost exponential in some
parameter (like the system size in exact diagonalization or some quantum
Monte Carlo methods)---or approximate, and therefore limited to certain
regimes (think of tensor networks, which are limited to weakly entangled
states).

These limitations (some of which were known when quantum computers
were first proposed) make processors with quantum properties---commonly
called quantum computers or quantum simulators---ideal candidates
for computing the spectral function: if these many-body systems can
be engineered or programmed to follow similar dynamics to Hubbard
dynamics, the time evolution of the processor will require resources
than scale, at least at face value, only linearly with the size of
the system and evolution time: to reach, say, larger lattice sizes,
one just has to add more ``particles'' (or quantum bits, as we shall
call them), and perform longer time evolutions... raising hope for
exponential speedups to perform time evolutions. Several quantum algorithms,
which we will explain in this lecture, were proposed to exploit this
fact. 

However, as we shall see, quantum algorithms are subject to strong
constraints inherent to their quantum nature. A major constraint,
which became obvious with the advent of physical realizations of quantum
computers in the last decade, is decoherence, namely unwanted entanglement
with the outside environment. Decoherence places hard limitations
on the duration (number of operations) of quantum algorithms, which
rules out many textbook quantum algorithms if no countermeasures are
taken. It can possibly be suppressed with quantum error correction
techniques, but these in turn require formidable resources that will
remain out of the reach of quantum processors for many years. Worse
still, even in the absence of decoherence, the local nature of available
operators and measurements, and the projective nature of the latter,
also need to be taken into consideration when designing a quantum
algorithm, and when comparing it to classical counterparts. Finally,
even on a complexity-theoretic level, preparing ground states or low-temperature
states of many-body systems is likely hard (that is, exponential)
even for perfect quantum computers.

Quantum processors are therefore not to be considered a silver bullets
to solve strongly-correlated models like the Hubbard model, but rather
as powerful heuristics that could outperform classical heuristics
in some difficult regimes whose precise delineations yet need to be
determined... One should perhaps even consider quantum processors
as coprocessors to be used in combination with classical heuristics
to reach regimes hitherto inaccessible to either classical or quantum
algorithms: in the same way as graphics processing units (GPUs) are
now used rountinely to speed up some linear algebra operations, quantum
processing units (QPUs) could be used to accelerate some well-defined
subroutines of an otherwise classical program.

\begin{figure}
\begin{centering}
\includegraphics[viewport=80bp 360bp 500bp 540bp,clip,width=0.8\columnwidth]{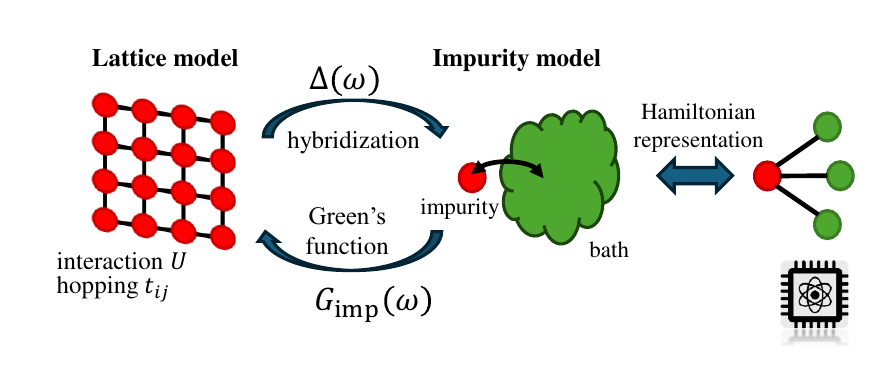}
\par\end{centering}
\caption{DMFT self-consistency cycle: a lattice model (here the Hubbard model)
is self-consistently mapped to an impurity model, defined by its hybridization
function $\Delta(\omega)$. This impurity model can be represented
with an Anderson impurity Hamiltonian describing a correlated impurity
site coupled to noninteracting bath sites (right). Quantum processors
can be used to solve the impurity model, namely compute its Green's
function $G_{\mathrm{imp}}(\omega)$.\label{fig:DMFT-self-consistency-cycle.}}
\end{figure}

\subsection{A method to reduce the complexity of the problem}

One very successful classical heuristic for tackling the Hubbard model
is dynamical mean field theory (DMFT, \cite{Georges1996}). It maps
the Hubbard model to a simpler, yet still many-body problem called
an impurity model. This impurity model describes one (or a few) interacting
fermionic sites embedded in a noninteracting environment. The properties
of this environment are adjusted to have the Green's function of the
impurity $G_{\mathrm{imp}}^{\mathrm{R}}(\omega)$ match the local
component of the lattice Green's function $G_{\mathrm{loc}}^{\mathrm{R}}(\omega)=\sum_{\boldsymbol{k}}G^{\mathrm{R}}(\boldsymbol{k},\omega)$.
This mapping is exact only in the limit of lattices of infinite dimensions
thanks to the local nature of Hubbard interactions, but DMFT is typically
used in two dimensions. Yet, it is able, among other successes, to
capture the Fermi-liquid to Mott-insulator transition. Perhaps more
importantly, DMFT comes with a control parameter, the number $N_{c}$
of correlated sites (also called the size of the impurity ``cluster'')
of the impurity model: it becomes exact in the limit of infinite $N_{c}$;
gradually increasing $N_{c}$ until convergence of quantitites of
interest provides a well-defined way of quantifying errors.

Over the years, very sophisticated methods to solve this correlated,
yet simpler impurity model, called ``impurity solvers'', have been
developed. While some of these impurity solvers (think of the numerical
renormalization group or segment-picture continuous-time quantum Monte
Carlo solvers) yield essentially exact solutions for the $N_{c}=1$
case at equilibrium, they face a number of difficulties beyond this
regime. For instance, as the number $N_{c}$ of impurities (or orbitals
when going beyond the single-orbital Hubbard model) increases, the
exponential wall plaguing classical methods to tackle the Hubbard
model makes its comeback. Besides, many of these solvers (like quantum
Monte-Carlo methods) work on the imaginary time axis, and thus require
uncontrolled analytical continuation techniques to obtain data on
the real axis. Conversely, most real-axis techniques (like exact diagonalization)
must deal with the large number of bath sites used to represent the
noninteracting environment, and the associated exponential complexity.
As for tensor network techniques, the time evolution required to obtain
Green's functions may lead to large entanglement levels and thus make
these methods either unreliable or unpractical. All these issues are
made worse when turning to out-of-equilibrium regimes, which are beyond
the scope of this lecture.

Given these limitations of impurity solvers, quantum computers, now
that the problem has been ``distilled'' down from a large lattice
problem to its ``quantum quintessence'', an impurity model, appear
as interesting candidates to reach regimes that are out of the reach
of classical impurity solvers: for instance, the regime of large cluster
size $N_{c}$, or the long-time limit needed to resolve low-energy
components of the spectral function. Even though the inherent difficulties
of quantum computers are still there, they are made less severe by
the reduction to an impurity model.

The goal of this lecture is to explain how the impurity problem of
DMFT could be tackled by quantum computers in theory (that is, with
perfect or error-corrected quantum computers), but also in practice
(taking into account the limitations of current and near-term quantum
hardware). To this aim, we will first very briefly recall the DMFT
formalism (illustrated in Fig. \ref{fig:DMFT-self-consistency-cycle.}),
and then introduce the basic tools of quantum computing. Then, we
will describe the textbook quantum algorithms that can be used to
solve impurity models, before turning to the practical issues and
how other types of algorithms can be used to try and overcome those
issues.

\section{Dynamical mean field theory and impurity solvers}

In this section, we briefly introduce the main DMFT concepts, and
in particular the main computational bottleneck of DMFT, the impurity
model. We also explain the main challenges of classical methods.

\subsection{Self-consistency equations}

The main target of equilibrium DMFT is the lattice's retarded Green's
function, $G^{\mathrm{R}}(\boldsymbol{k},\omega)$ (Eq. (\ref{eq:retarded_gf})).
A perturbation expansion of the Hubbard Hamiltonian (Eq. (\ref{eq:hubbard_model}))
in powers of the interaction $U$ generates a series expansion around
the unperturbed Green's function $G_{0}^{\mathrm{R}}(\boldsymbol{k},\omega)$.
This series can in turns be cleverly resummed to yield the following
equation, called the Dyson equation:
\begin{equation}
G^{\mathrm{R}}(\boldsymbol{k},\omega)=G_{0}^{\mathrm{R}}(\boldsymbol{k},\omega)+G_{0}^{\mathrm{R}}(\boldsymbol{k},\omega)\Sigma(\boldsymbol{k},\omega)G^{\mathrm{R}}(\boldsymbol{k},\omega),\label{eq:Dyson}
\end{equation}
with a new object, $\Sigma(\boldsymbol{k},\omega)$, called the self-energy,
that vanishes when $U$ vanishes. This self-energy captures the effect
of interactions on the propagation of electrons. In the limit when
the Hubbard model is defined on a lattice with infinite dimensions,
the self-energy becomes local, namely independent of $\boldsymbol{k}$.
DMFT consists in (i) making the approximation that the self-energy
remains local also in finite (and even low) dimensions, and (ii) using
a surrogate model, which we will later call impurity model, from which
a local self-energy $\Sigma_{\mathrm{imp}}(\omega)$ can be computed
as an approximation to the lattice self-energy. This surrogate model
also originates from the infinite-dimensional limit of the Hubbard
model. It is defined by its action
\begin{equation}
S_{\mathrm{imp}}=\iint_{0}^{\beta}\mathrm{d}\tau\mathrm{d}\tau'\sum_{\sigma}c_{\sigma}^{*}(\tau)\left(-\mathcal{G}_{0}^{-1}(\tau-\tau')\right)c_{\sigma}(\tau')+\int_{0}^{\beta}\mathrm{d}\tau Un_{\uparrow}(\tau)n_{\downarrow}(\tau).\label{eq:impurity_action}
\end{equation}
Here, $c_{i\sigma}^{*}(\tau)$ and $c_{i\sigma}(\tau)$ denote two
Grassmann fields. $S_{\mathrm{imp}}$ depends entirely on $U$ and
on the noninteracting Green's function $\mathcal{G}_{0}$ (here given
in imaginary time $\tau$ for simplicity; in general, $f^{\mathrm{R}}(\omega)$
can be recovered, at least formally, from $f(\tau)$ by first Fourier
transforming $f(\tau)$ to $\tilde{f}(i\omega_{n})=\int_{0}^{\beta}\mathrm{d}\tau e^{i\tau\omega_{n}}f(\tau)$
and then performing an analytical continuation $f^{\mathrm{R}}(\omega)=\tilde{f}(z=\omega+i\eta)$).
$\mathcal{G}_{0}^{\mathrm{R}}$, $\Sigma_{\mathrm{imp}}^{\mathrm{R}}$
and the impurity Green's function $G_{\mathrm{imp}}^{\mathrm{R}}$
are related by Dyson's equation: $\Sigma_{\mathrm{imp}}^{\mathrm{R}}(\omega)=\mathcal{G}_{0}^{-1}(\omega)-(G_{\mathrm{imp}}^{\mathrm{R}})^{-1}(\omega)$.
One often defines the hybridization function $\Delta^{\mathrm{R}}(\omega)$
as $\Delta^{\mathrm{R}}(\omega)=\omega+\mu-\mathcal{G}_{0}^{-1}(\omega)$.
With this, the impurity action reads 
\begin{equation}
S_{\mathrm{imp}}=S_{\mathrm{loc}}+S_{\mathrm{hyb}},\label{eq:impurity_action-hyb}
\end{equation}
Here, $S_{\mathrm{loc}}=\int_{0}^{\beta}\mathrm{d}\tau\sum_{\sigma}c_{\sigma}^{*}(\tau)\left(\partial_{\tau}-\mu\right)c_{\sigma}(\tau)+\int_{0}^{\beta}\mathrm{d}\tau Un_{\uparrow}(\tau)n_{\downarrow}(\tau)$
is easily seen to be the action of a single correlated site, namely
a single fermionic site with potential energy $-\mu$ and interaction
energy $U$ to penalize double occupancies. The hybridization term
$S_{\mathrm{hyb}}=\iint_{0}^{\beta}\mathrm{d}\tau\mathrm{d}\tau'\sum_{\sigma}c_{\sigma}^{*}(\tau)\Delta(\tau-\tau')c_{\sigma}(\tau')$
in $S_{\mathrm{imp}}$ describes how this site is coupled to an environment
that is completely characterized by a (dynamical) mean field $\Delta^{\mathrm{R}}(\omega)$.
This dynamical mean field a priori describes an infinite number of
degrees of freedom. Since the spectral function is the central object
of interest, one adjusts the surrogate model (and hence $\Delta^{\mathrm{R}}(\omega)$)
in such a way that its Green's function $G_{\mathrm{imp}}^{\mathrm{R}}(\omega)$
coincides with the local component of $G^{\mathrm{R}}(\boldsymbol{k},\omega)$:
\begin{equation}
G_{\mathrm{imp}}^{\mathrm{R}}(\omega)\left[\Delta^{\mathrm{R}}\right]=\sum_{\boldsymbol{k}}G^{\mathrm{R}}(\boldsymbol{k},\omega)\left[\Delta^{\mathrm{R}}\right].\label{eq:dmft_self_consistency}
\end{equation}
In this self-consistency equation, we made explicit the functional
dependence on $\Delta^{\mathrm{R}}$: (i) as the Green's function
of $S_{\mathrm{imp}}$, $G_{\mathrm{imp}}^{\mathrm{R}}$ is directly
a functional of $\Delta^{\mathrm{R}}$ (see Eq. (\ref{eq:g_imp_grassmann})
below); (ii) the dependence of $G^{\mathrm{R}}(\boldsymbol{k},\omega)$
comes from Dyson's equation (\ref{eq:Dyson}), which, when performing
the DMFT approximation $\Sigma(\boldsymbol{k},\omega)\approx\Sigma_{\mathrm{imp}}(\omega)$,
explicitly reads

\begin{equation}
G^{\mathrm{R}}(\boldsymbol{k},\omega)=\frac{1}{\omega-\varepsilon(\boldsymbol{k})+\mu-\Sigma_{\mathrm{imp}}^{\mathrm{R}}(\omega)[\Delta^{\mathrm{R}}]},\label{eq:lattice_gf}
\end{equation}
with $\varepsilon(\boldsymbol{k})$ the space Fourier transform of
the hopping matrix $t_{ij}$ of the Hubbard model (Eq. (\ref{eq:hubbard_model})).
Finally, $\Sigma_{\mathrm{imp}}^{\mathrm{R}}(\omega)[\Delta^{\mathrm{R}}]=\omega+\mu-\Delta^{\mathrm{R}}-G_{\mathrm{imp}}^{\mathrm{R}}(\omega)^{-1}\left[\Delta^{\mathrm{R}}\right]$.
Putting everything together, we obtain a fixed-point equation:
\begin{equation}
G_{\mathrm{imp}}^{\mathrm{R}}(\omega)\left[\Delta^{\mathrm{R}}\right]=\sum_{\boldsymbol{k}}\frac{1}{\Delta^{\mathrm{R}}(\omega)+G_{\mathrm{imp}}^{\mathrm{R}}(\omega)^{-1}\left[\Delta^{\mathrm{R}}\right]-\varepsilon(\boldsymbol{k})}.\label{eq:self-consistency-final}
\end{equation}
Solving DMFT amounts to adjusting $\Delta^{\mathrm{R}}(\omega)$ to
fulfill the above equation. It is usually solved iteratively. The
process crucially hinges on the ability to find $G_{\mathrm{imp}}^{\mathrm{R}}(\omega)$
for a given $\Delta^{\mathrm{R}}(\omega)$. We will henceforth refer
to this task as ``solving the impurity model'': it is the bottleneck
of DMFT.

We note that the derivation above can easily be adapted to the case
of several correlated atoms (or ``impurities'') instead of one (this
is then called ``cluster DMFT''), and to the nonequilibrium case
(``out-of-equilibrium DMFT'').

Before turning to solving impurity models, let us emphasize that imposing
constraints at the level of the (single-particle) Green's function
is a choice driven by the physical question at stake (here the study
of phase transitions with order parameters related to $G$). Other
choices are possible, leading to a whole spectrum of so-called quantum
embedding theories (of which DMFT is the earliest representative):
one may require self-consistency on ``simpler'' objects (like one-particle
reduced density matrices $\langle c_{i}^{\dagger}c_{j}\rangle$ (same
as DMFT but with no time dependence), leading to theories like density-matrix
embedding theory (DMET, \cite{Knizia2012}) or rotationally-invariant
slave bosons (RISB, \cite{Lechermann2007,Ayral2017a}))), or on more
sophisticated objects (like two-particle Green's functions, leading
to theories like the dynamical vertex approximation, D$\Gamma$A\cite{Held2014}).
In the former case, one can have the intuition that the corresponding
surrogate model will be simpler; conversely, in the latter case, the
impurity model should be more complicated \cite{Ayral2016}). 

\subsection{Classical impurity solvers... and their limitations}

In this section, we briefly address the problem of finding the Green's
function $G_{\mathrm{imp}}$ for a given hybridization function $\Delta$.
In a Grassmann path integral formalism, it is given, as a function
of imaginary time $\tau$, by the expression:
\begin{equation}
G_{\mathrm{imp}}(\tau)=-\int\mathcal{D}\left[c,c^{*}\right]c_{\sigma}(\tau)c_{\sigma}^{*}(0)e^{-S_{\mathrm{imp}}\left[\Delta\right]}.\label{eq:g_imp_grassmann}
\end{equation}
(We dropped the $\sigma$ dependence in $G_{\mathrm{imp}}$ to simplify
notation, but phases with spin-symmetry breaking can be studied).
Impurity solvers can be classified in two main families, Hamiltonian-based
solvers and action-based solvers.

\subsubsection{Hamiltonian-based solvers}

We start with Hamiltonian-based solvers, whose formalism is more directly
translatable to quantum algorithms. In Hamiltonian-based solvers,
one introduces a Hamiltonian model, dubbed the ``Anderson impurity
model'' (AIM)\index{Anderson impurity model},
\begin{equation}
H_{\mathrm{AIM}}=Un_{\uparrow}n_{\downarrow}-\mu\sum_{\sigma}c_{\sigma}^{\dagger}c_{\sigma}+\sum_{k\sigma}\epsilon_{k}a_{k\sigma}^{\dagger}a_{k\sigma}+\sum_{k\sigma}V_{k}\left(a_{k\sigma}^{\dagger}c_{\sigma}+\mathrm{h.c}\right),\label{eq:AIM_hamiltonian}
\end{equation}
which describes a single atom (called the impurity) with potential
energy $-\mu$ and interaction energy $U$ coupled via hopping terms
$V_{k}$ to a bath of noninteracting fermions with energies $\epsilon_{k}$
(and creation (resp. annihilation) operators $a_{k\sigma}^{\dagger}$
(resp. $a_{k\sigma}$)). This model was originally introduced by Anderson
to describe isolated impurities in metals \cite{Anderson1961}, hence
its name. It is also central to explain so-called Kondo physics (a
Schrieffer-Wolff transformation of this model yields its low-energy
simplification, the Kondo model). We are now going to show that the
hopping $V_{k}$ and energy $\epsilon_{k}$ parameters can be adjusted
so that the Green's function of this model coincides with the impurity
Green's function (Eq. (\ref{eq:g_imp_grassmann})).

The Green's function of the impurity site in $H_{\mathrm{AIM}}$ is
given by
\begin{equation}
G_{\mathrm{AIM}}(\tau)=-\int\mathcal{D}\left[c,c^{*},a_{k},a_{k}^{*}\right]c_{\sigma}(\tau)c_{\sigma}^{*}(0)e^{-S_{\mathrm{AIM}}},\label{eq:G_AIM}
\end{equation}
where $S_{\mathrm{AIM}}$ is the action corresponding to $H_{\mathrm{AIM}}$.
Since $S_{\mathrm{AIM}}$ is quadratic in the bath fields $a_{k}(\tau)$
and $a_{k}(\tau)^{*}$, they can be integrated out:
\[
\int\mathcal{D}\left[a_{k},a_{k}^{*}\right]e^{-S_{\mathrm{AIM}}}=e^{-S_{\mathrm{loc}}-\iint_{0}^{\beta}\mathrm{d}\tau\mathrm{d}\tau'\sum_{\sigma}c_{\sigma}^{*}(\tau)\Delta_{\mathrm{AIM}}(\tau-\tau')c_{\sigma}(\tau')}
\]
with $\Delta_{\mathrm{AIM}}^{\mathrm{R}}(\omega)=\sum_{k}\frac{V_{k}^{2}}{\omega+i\eta-\epsilon_{k}}.$
By comparing Eqs (\ref{eq:G_AIM}) and (\ref{eq:g_imp_grassmann}),
we see that $G_{\mathrm{AIM}}$ can be made to coincide with $G_{\mathrm{imp}}$
provided $\Delta_{\mathrm{AIM}}^{\mathrm{R}}(\omega)=\Delta^{\mathrm{R}}(\omega)$.
In other words, if one can find parameters $\left(V_{k},\epsilon_{k}\right)$
such that
\begin{equation}
\Delta^{\mathrm{R}}(\omega)=\sum_{k}\frac{V_{k}^{2}}{\omega+i\eta-\epsilon_{k}},\label{eq:fit_problem}
\end{equation}
then the AIM's Green's function coincides with the impurity's. This
exact correspondance (up to a fit) between our action-based impurity
model and a Hamiltonian model makes it possible to tackle the impurity
model with a variety of Hamiltonian-based techniques. Before we turn
to the main methods and their limitations, let us emphasize that the
fit problem (Eq. (\ref{eq:fit_problem})) itself is not straighforward:
in practice, one needs to perform this fit with a finite number of
bath orbitals, possibly leading to fitting errors and finite-size
effects. Different choice of metrics to optimize the fitting parameters
$(V_{k},\epsilon_{k})$ will a priori lead to different results.

Hamiltonian-based techniques consist in finding the eigenvectors $|\Psi_{\alpha}\rangle$
and eigenvalues $E_{\alpha}$ of $H_{\mathrm{AIM}}$ to compute the
Green's function. Indeed, one can rewrite, in operator formalism,
$G_{\mathrm{AIM}}(\tau)=-\mathrm{Tr}\left[\rho c_{\sigma}(\tau)c^{\dagger}(0)\right],$with
$\rho=e^{-\beta H_{\mathrm{AIM}}}/Z$ (and $\tau>0$ here). Inserting
completeness relations and expanding the trace, we get $G_{\mathrm{AIM}}(\tau)=-\frac{1}{Z}\sum_{\alpha,\alpha'}e^{-\beta E_{\alpha}}e^{-\tau(E_{\alpha'}-E_{\alpha})}\left|\langle\Psi_{\alpha}|c|\Psi_{\alpha'}\rangle\right|^{2}$
that becomes, once Fourier-transformed to Matsubara frequencies $i\omega_{n}$,
the so-called Lehmann representation
\begin{equation}
G_{\mathrm{AIM}}(i\omega_{n})=-\frac{1}{Z}\sum_{\alpha,\alpha'}\frac{e^{-\beta E_{\alpha'}}+e^{-\beta E_{\alpha}}}{i\omega_{n}-(E_{\alpha}-E_{\alpha'})}\left|\langle\Psi_{\alpha}|c|\Psi_{\alpha'}\rangle\right|^{2}.\label{eq:Lehmann_freq}
\end{equation}
In the zero-temperature limit ($\beta\rightarrow\infty$), 
\begin{equation}
G_{\mathrm{AIM}}(i\omega_{n})=-\sum_{\alpha\in\mathcal{H}_{N_{e}+1}}\frac{1}{i\omega_{n}-(E_{0}-E_{\alpha})}\left|\langle\Psi_{0}|c|\Psi_{\alpha}\rangle\right|^{2}-\sum_{\alpha\in\mathcal{H}_{N_{e}-1}}\frac{1}{i\omega_{n}-(E_{\alpha}-E_{0})}\left|\langle\Psi_{\alpha}|c|\Psi_{0}\rangle\right|^{2}.\label{eq:Lehmann_zero_temp}
\end{equation}
where we supposed that the ground state $|\Psi_{0}\rangle$ contains
$N_{e}$ electrons, and $\mathcal{H}_{N_{e}\pm1}$ denotes the eigenspace
with $N_{e}\pm1$ electrons. The spectral function $A_{\mathrm{imp}}(\omega)=-\frac{1}{\pi}\mathrm{Im}G_{\mathrm{AIM}}(z=\omega+i\eta)$
thus has Dirac delta peaks at energies $\omega=E_{0}-E_{\alpha}$
(with $\alpha\in\mathcal{H}_{N_{e}+1}$, namely minus the electron
addition energy) and $\omega=E_{\alpha}-E_{0}$ (with $\alpha\in\mathcal{H}_{N_{e}-1}$,
namely the electron removal energy).

The numerical challenge thus consists in finding the eigenvectors
$|\Psi_{\alpha}\rangle$ and eigenvalues $E_{\alpha}$ of $H_{\mathrm{AIM}}$
(at low temperatures, only the low-lying ones). The sheer size of
the Hilbert space ($4^{N_{\mathrm{c}}+N_{\mathrm{b}}}$ if we denote
by $N_{\mathrm{b}}$ the number of bath sites) makes it a difficult
problem, but the methods below either try to exploit the sparsity
of $H_{\mathrm{AIM}}$ and the fact that only the low-lying states
are needed, or use a compressed state representation that works well
when entanglement is low.

\paragraph{Exact diagonalization: Lanczos method.\label{par:Exact-diagonalization:-Lanczos}}

A first method, called the Lanczos method (see \cite{Koch2011}),
avoids the $O\left(\left(4^{N_{c}+N_{b}}\right)^{3}\right)$ cost
of a direct diagonalization of $H_{\mathrm{AIM}}$ by using the sparsity
of $H_{\mathrm{AIM}}$: in the Fock basis representation of $H_{\mathrm{AIM}}$
(Eq. (\ref{eq:AIM_hamiltonian})), the number of nonzero elements
per row or column of $H_{\mathrm{AIM}}$ is of the order of $s=O(N_{c}+N_{b})$.
Thus, matrix-vector multiplications $H_{\mathrm{AIM}}|\Psi\rangle$
take only $O\left(s4^{N_{c}+N_{b}}\right)$ time. This property is
exploited in the Lanczos method by finding the matrix of $H_{\mathrm{AIM}}$
in the so called Krylov basis $\left\{ |\chi_{n}\rangle\right\} _{n=1\dots K}$
of the $K$-dimensional vector space spanned by the family $\left\{ |\Phi_{0}\rangle,H_{\mathrm{AIM}}|\Phi_{0}\rangle,H_{\mathrm{AIM}}^{2}|\Phi_{0}\rangle,\dots,H_{\mathrm{AIM}}^{K}|\Phi_{0}\rangle\right\} $,
with $|\Phi_{0}\rangle$ is an initial state that must not be orthogonal
to the exact ground state $|\Psi_{0}\rangle$. $H_{\mathrm{AIM}}$
is tridiagonal in this basis, which makes it easy to compute its ground
state energy. The error on the ground state energy obtained in this
reduced space decreases exponentially with $K$~\cite{Koch2011}.

Let us now turn to the Green's function. Using Eq. (\ref{eq:retarded_gf})
in the case of the AIM and at zero temperature, we obtain:

\begin{align}
G^{\mathrm{R}}(t) & =\theta(t)\left(-ie^{iE_{0}t}\langle\Psi_{0}|c_{\sigma}e^{-iH_{\mathrm{AIM}}t}c_{\sigma}^{\dagger}|\Psi_{0}\rangle-ie^{-iE_{0}t}\langle\Psi_{0}|c_{\sigma}^{\dagger}e^{iH_{\mathrm{AIM}}t}c_{\sigma}|\Psi_{0}\rangle\right).\label{eq:G_R_AIM}
\end{align}
Taking the Fourier transform of the first term (the second term, $G^{<}(\omega)$,
can be dealt with in a similar fashion), and inserting resolutions
of the identity, we obtain
\begin{align}
G^{>}(\omega) & =-i\sum_{\alpha}\int_{0}^{\infty}\mathrm{d}te^{i\omega t}\langle\Psi_{0}|e^{iH_{\mathrm{AIM}}t}c_{\sigma}e^{-iH_{\mathrm{AIM}}t}|\Psi_{\alpha}\rangle\langle\Psi_{\alpha}|c_{\sigma}^{\dagger}|\Psi_{0}\rangle\label{eq:G_greater_freq-1}\\
 & =-i\sum_{\alpha}\int_{0}^{\infty}\mathrm{d}te^{i(\omega+E_{0}-E_{\alpha})t}\left|\langle\Psi_{\alpha}|c_{\sigma}^{\dagger}|\Psi_{0}\rangle\right|^{2}.\label{eq:G_greater_2}
\end{align}
Performing the integration and adding a $\eta>0$ factor to ensure
convergence, we find the so-called resolvent form of the Green's function:
\begin{align}
G^{>}(\omega+i\eta) & =\sum_{\alpha}\frac{\left|\langle\Psi_{0}|c_{\sigma}|\Psi_{\alpha}\rangle\right|^{2}}{E_{0}-E_{\alpha}+\omega+i\eta}=\langle\Psi_{0}|c_{\sigma}\frac{1}{E_{0}-H+\omega+i\eta}c_{\sigma}^{\dagger}|\Psi_{0}\rangle.\label{eq:resolvent_g_greater}
\end{align}
Having computed $|\Psi_{0}\rangle$ and $E_{0}$ in the first Lanczos
step, one can now perform a second Lanczos step, this time with $|\Phi_{0}\rangle\propto c_{\sigma}^{\dagger}|\Psi_{0}\rangle$.
In the corresponding second Krylov basis, taking advantage of the
fact that $H$ is tridiagonal in any Krylov basis, one can easily
write a continued fraction expression for $G^{>}(\omega+i\eta)$.
One can thus obtain an approximation of the Green's function in real
frequency with a run time that scales (with a naive estimate) as $O(sK4^{N_{c}+N_{b}})$. 

\paragraph{Matrix product states: density matrix renormalization group method\label{par:Matrix-product-states:}}

\begin{figure}
\begin{centering}
\includegraphics[width=0.9\columnwidth]{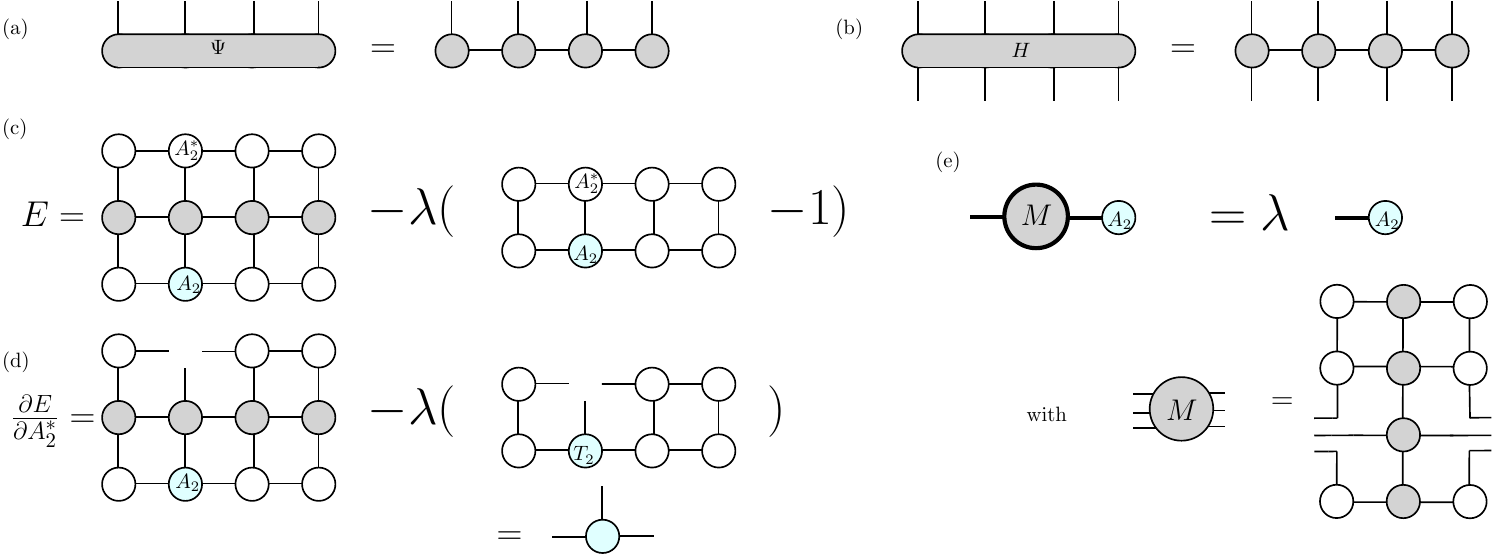}
\par\end{centering}
\caption{Matrix product states and the density matrix renormalization group
method. (a) MPS representation of $|\Psi\rangle$. (b) MPO representation
of $H$. (c) Tensor-network representation of the energy functional
of DMRG. (d) TN representation of a partial derivative. (e) TN representation
of the eigenproblem (thick lines denote groupings of tensor legs).\label{fig:Matrix-product-states}}

\end{figure}

A second method allows avoiding the exponential cost of storing the
wave function by assuming the bipartite entanglement in the system
is low. Instead of representing the wavefunction as
\begin{equation}
|\Psi\rangle=\sum_{b_{1},\dots b_{N}}\psi_{b_{1},\dots b_{N}}|b_{1},\dots,b_{N}\rangle,\label{eq:psi_n_qubits}
\end{equation}
with $N=2\left(N_{c}+N_{\mathrm{b}}\right)$, with associated storage
cost $2^{N}$, we assume the wavefunction amplitudes to be representable
as
\begin{equation}
\psi_{b_{1},\dots b_{N}}=\sum_{\alpha_{1},\dots\alpha_{N-1}}\left[A^{(1)}\right]_{\alpha_{1}}^{b_{1}}\left[A^{(2)}\right]_{\alpha_{1},\alpha_{2}}^{b_{2}}\cdots\left[A^{(2)}\right]_{\alpha_{N-2},\alpha_{N-1}}^{b_{N-1}}\left[A^{(1)}\right]_{\alpha_{N-1}}^{b_{N}},\label{eq:MPS_form}
\end{equation}
with $N$ rank-3 tensors $\left[A^{(k)}\right]_{\alpha_{k-1},\alpha_{k}}^{b_{k}}$
---or matrices if one considers $b_{k}$ to be fixed, hence the name
``matrix product state'' (MPS) representation\index{matrix product states}.
This is illustrated in Fig. \ref{fig:Matrix-product-states} (a).
Each tensor index has dimension $2$, $\chi$, $\chi$ for $b_{k}$,
$\alpha_{k-1}$ and $\alpha_{k}$, respectively, with $\chi$ called
the bond dimension of the MPS. The total storage cost thus scales
as $O(N\chi^{2})$: it is now linear in the number $N\propto N_{\mathrm{c}}+N_{\mathrm{b}}$
of sites! However, a fixed bond dimension $\chi$ places a limitation
on how large the so-called von Neumann entanglement entropy $S$ of
the state can be. One can easily show that $\chi$ needs to fulfill
$\chi\geq2^{S}$ (\cite{Schollwoeck2011}); otherwise, truncation
errors appear. It turns out that for Anderson impurity models with
$N_{\mathrm{c}}=1$, one can always make the problem one-dimensional,
in which case known results about the entanglement entropy of ground
states of one-dimensional Hamiltonians apply: $S$ follows an ``area
law'', which in 1D means that $S\propto\mathrm{const}$. This allows
for a very inexpensive storage of the ground state of the AIM for
the $N_{\mathrm{c}}=1$ case.

To find the actual MPS form of the ground state, one usually resorts
to a variational algorithm known as the density matrix renormalization
group algorithm (DMRG), which consists in minimizing $\langle\Psi(A^{(1)},\dots A^{(N)})|H|\Psi(A^{(1)},\dots,A^{(N)})\rangle$
while imposing $\langle\Psi(A^{(1)},\dots A^{(N)})|\Psi(A^{(1)},\dots,A^{(N)})\rangle=1$
(Fig. \ref{fig:Matrix-product-states} (c)). Writing stationarity
conditions for a given tensor $A^{(k)}$ (given a fixed value of the
other $N-1$ tensors) leads to an eigenvalue problem $H_{\mathrm{eff}}^{(k)}\psi^{(k)}=E^{(k)}\psi^{(k)}$
with $\psi^{(k)}$ of size $2\chi^{2}$ (Fig. \ref{fig:Matrix-product-states}
(e)): it can be easily solved exactly. The process is then iterated
over $k=1\dots N$ several times (sweeps) until convergence of the
individual tensors. The computational cost thus scales as $N_{\mathrm{s}}\left(2\chi^{2}\right)^{3}$,
with $N_{\mathrm{s}}$ the number of sweeps, if one uses a direct
diagonalization procedure.

With the ground state in hand, one can now in principle compute the
Green's function: in Eq. (\ref{eq:G_R_AIM}), we see that to get the
first term, we need to start from $|\Psi_{0}\rangle$, apply a creation
operator to get $c_{\sigma}^{\dagger}|\Psi_{0}\rangle$, time evolve
($e^{-iH_{\mathrm{AIM}}t}$) and then project back onto $c_{\sigma}^{\dagger}|\Psi_{0}\rangle$.
The operation $c_{\sigma}^{\dagger}|\Psi_{0}\rangle$ is straightforward
to perform with matrix product states. The time evolution can be performed
with a number of methods. The most straightforward one is time-evolving
block decimation (TEBD), which consists in splitting the time-evolution
operator $e^{-iH_{\mathrm{AIM}}t}$ in a sequence of so-called Trotter
time steps,
\begin{equation}
e^{-iH_{\mathrm{AIM}}t}=\left(e^{-iH_{\mathrm{AIM}}t/N_{\mathrm{t}}}\right)^{N_{\mathrm{t}}}\approx\left(1-iH_{\mathrm{AIM}}\frac{t}{N_{\mathrm{t}}}\right)^{N_{\mathrm{t}}}.\label{eq:TEBD_trotter}
\end{equation}
Then, the operator $\delta U\equiv1-iH_{\mathrm{AIM}}\frac{t}{N_{\mathrm{t}}}$
(represented in so-called matrix product operator, MPO, format) is
applied to the current wavevector (represented in MPS format) $N_{\mathrm{t}}$
times. This is in principle simple, but the bond dimension $\chi$
of the MPS is not conserved upon application of $\delta U$: it generically
increases, because time evolving the ground state generates excited
states that are usually more entangled than the ground state, and
thus requires a larger bond dimension to be represented exactly. One
faces a dilemma: either ones tries to keep up with the increasing
entanglement by increasing $\chi$ (but one quickly reaches memory
limits), or one truncates the MPS to a fixed maximal bond dimension,
thereby incurring truncation errors. More sophisticated methods exists
(see \cite{Paeckel2019} for a review), but they generically meet
the challenge of an increasing entanglement entropy with evolution
time $t$.

\subsubsection{Action-based solvers\label{subsec:Action-based-solvers}}

The above methods rely on a Hamiltonian representation (Eq. (\ref{eq:AIM_hamiltonian}))
of the impurity action (\ref{eq:impurity_action-hyb}). A major drawback
of these methods is that they require a Hamiltonian model with a large
bath (large $N_{\mathrm{b}}$) to faithfully represent the hybridization
function of DMFT. This large bath, in turn, leads to large numerical
costs (unless one chooses a small bath... but then suffers from large
finite-size effects).

Action-based solvers avoid this problem by working directly in the
action formalism. Starting from Eq. (\ref{eq:impurity_action-hyb}),
they fall into two main categories, depending on whether they use
an expansion in powers of $S_{\mathrm{loc}}$ (interaction expansion
solvers) or $S_{\mathrm{hyb}}$ (hybridization expansion solvers).
The resulting infinite sum is sampled using Monte-Carlo methods. The
two categories come with distrinct properties: intuitively, interaction
expansion solvers are better behaved when interactions are weak, while
hybridization expansion solvers are better behaved in the strong interaction
limit. Both families yield the impurity Green's function in imaginary
time $G\mathrm{_{imp}}(\tau)$ exactly up to statistical uncertainty.
It turns out that this statistical uncertainty (namely the variance
of the Monte-Carlo averages used to compute the Green's function)
generically blows up (exponentially) with decreasing temperature and
with increasing cluster size (or number of orbitals) $N_{\mathrm{c}}$,
a manifestation of the so-called fermionic sign problem. To curb this
variance, one in principle needs very long Markov chains and thus
very long run times.

Let us also stress that even in the absence of this sign problem (say
in the $N_{c}=1$ case), obtaining the real-time frequency Green's
function $G_{\mathrm{imp}}^{\mathrm{R}}(\omega)$ from noisy (due
to Monte-Carlo statistical noise), imaginary-time $G_{\mathrm{imp}}(\tau)$,
a problem called the analytical continuation, is notoriously difficult
(it amounts to inverting an ill-conditioned matrix), so that it would
be highly desirable to work directly in real time. However, in real
time, Monte-Carlo methods struggle with another sign problem stemming
from the complex imaginary factors of the time evolution operator
(``dynamical sign problem''), strongly limiting these methods to
short times, and thus not allowing enough low-frequency accuracy in
the resulting spectral functions.

\section{Quantum computing tools for impurity models}

In the previous section, we saw that DMFT can be seen as a classical
heuristic that self-consistently reduces strongly correlated lattice
models like the Hubbard model to simpler, yet still strongly-correlated
models known as impurity models (Fig. \ref{fig:DMFT-self-consistency-cycle.}).
To compute the spectral function of the lattice model within the DMFT
approximation, one needs to compute the impurity retarded Green's
function, namely the Green's function of a few ($N_{\mathrm{c}}$)
correlated sites exchanging electrons with a large noninteracting
fermionic bath. Essentially, as we saw, the computation of this Green's
function boils down to computing objects of the form
\begin{equation}
G^{>}(t)=-i\langle c_{\sigma}(t)c_{\sigma}^{\dagger}\rangle\label{eq:G_greater}
\end{equation}
which, together with its counterpart $G^{<}(t)=i\langle c_{\sigma}^{\dagger}c_{\sigma}(t)\rangle$,
appear in the definition (\ref{eq:retarded_gf}) of the retarded Green's
function (here for simplicity we dropped the $\sigma$ dependence
of $G$, and assumed $N_{\mathrm{c}}=1$; also, because of the $\theta(t)$
appearing in (\ref{eq:retarded_gf}), we can assume $t>0$).

The goal of this section is to show how quantum computers can be used
to compute objects of form of $G^{>}(t)$ for the specific case of
impurity models. If we expand its zero-temperature expression, we
recall that we obtain
\begin{equation}
G^{>}(t)=-i\langle\Psi_{0}|e^{iH_{\mathrm{AIM}}t}c_{\sigma}e^{-iH_{\mathrm{AIM}}t}c_{\sigma}^{\dagger}|\Psi_{0}\rangle.\label{eq:G_greater_explicit}
\end{equation}
This form makes it clear that we can decompose the computation in
two parts: the computation of the ground state $|\Psi_{0}\rangle$,
and the time evolution $e^{-iH_{\mathrm{AIM}}t}$ of the ground state
with one added particle. This is similar to the strategy we described
above for matrix product states, but as we shall see, using a quantum
processor comes, at least in theory, with added benefits.

\subsection{Quantum computing in a nutshell\label{subsec:Quantum-computing-in}}

In this section, we introduce the main tools available to a quantum
programmer, with a focus on tools that will be useful for solving
impurity models. In particular, a goal of this section is to describe
a simple way to perform the operation $e^{-iH_{\mathrm{AIM}}t}$ on
a quantum computer.

\subsubsection{Definition of a quantum computer}

A quantum computer is a system described by a time-dependent Schr�dinger
equation
\begin{equation}
i\hbar\frac{\mathrm{d}|\Psi(t)\rangle}{\mathrm{d}t}=H(t)|\Psi(t)\rangle,\label{eq:schro}
\end{equation}
and whose initial state $|\Psi(t=0)\rangle$ and time-dependent Hamiltonian
$H(t)$ can be \emph{controlled} to reach a target final state $|\Psi(t_{\mathrm{f}})\rangle$,
on which observables (hermitian operators) $\hat{O}$ of interest
can be measured. This can be done either in a one-shot way (giving
access to a bit of information $\lambda$ with a probability given
by the overlap of $|\Psi(t_{\mathrm{f}})\rangle$ with one of the
eigenvectors $|\varphi_{\lambda}\rangle$ of $\hat{O}$), or in an
average way (giving access to a statistical estimate of the average
value $\langle\hat{O}\rangle=\langle\Psi(t_{\mathrm{f}})|\hat{O}|\Psi(t_{\mathrm{f}})\rangle$).

What this system is varies from one quantum computer implementation
to the other. A distinction is usually made between analog computers
(also known as quantum simulators)---whose Hamiltonian is very specific
(e.g with a limited control on the individual degrees of freedom),
and digital (or gate-based) computers---which are characterized by
the possibility to control individually each degree of freedom, with
a practical consequence: this individual (or local) control, complemented
with an entangling operation between the degrees of freedom, leads
to a form of ``universality'', namely any unitary operation $U$
can be approximated by the unitary operator corresponding to the full
time evolution,
\begin{equation}
U(t)=T\mathrm{exp}\left(-i\int_{0}^{t}H(\tau)\mathrm{d}\tau\right).\label{eq:U_dyson_series}
\end{equation}
A universal quantum computer can thus be regarded as a machine that
implements arbitrary unitary operations and measurements. The space on which these
operations are performed is, for mainstream quantum computers, the
Hilbert space of $N$ two-level systems (aka quantum bits or qubits)
$\mathcal{H}_{N}$. Its size is exponential ($\mathrm{dim}\mathcal{H}_{N}=2^{N}$),
and a generic $N$-qubit state (recall Eq. (\ref{eq:psi_n_qubits}))
a priori requires $2^{N}$ complex coefficients to be represented
on a classical computer. While simulating such a time evolution on
a classical computer would a priori involve resources scaling as $O(2^{N})$,
letting the quantum computer evolve ``naturally'' leads to resources
linear in the number of degrees of freedom $N$.

Let us note that quantum computers are not limited to qubit (or spin-1/2)
systems. In particular, a natural candidate implementation for dealing
with fermionic systems would be a system with fermionic degrees of
freedom like ultracold fermionic atoms in optical lattices, whose
Hamiltonian is engineered as close as possible to, say, the Hubbard
model. These systems are indeed promising platforms for gaining insights,
among others, into Hubbard physics, with very recent major improvements
in the reachable temperatures. In this lecture, however, we will
focus only on qubit-based computers, and within this category, on
gate-based quantum computers. As we shall see, using these computers
to tackle fermionic systems will require a translation from the world
of fermions to the world of qubits; but the exquisite control afforded
by this type of computers makes them very convenient to design algorithms.

\subsubsection{The circuit model}

\begin{figure}
\noindent \begin{centering}
\includegraphics[width=0.6\columnwidth]{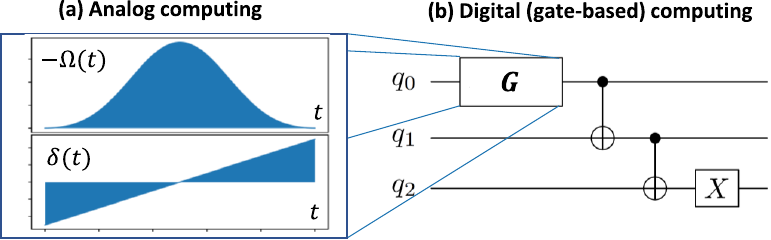}
\par\end{centering}
\caption{Circuit model: a quantum circuit (on the right) describes a sequence
of gates, which are individually implemented in an analog fashing
with time-dependent terms in the hardware Hamiltonian. Reproduced
from \cite{Ayral2023}.\label{fig:Circuit-model.}}
\end{figure}

One usually picks, as a basis of the Hilbert space $\mathcal{H}_{n}$,
the tensor basis of the eigenstates of Pauli-z matrix $\sigma_{z}$:
it is denoted as $|b_{1},\dots,b_{N}\rangle$, with $b_{k}\in\left\{ 0,1\right\} $
and $\sigma_{z}^{(k)}|b_{k}\rangle=(-)^{b_{k}}|b_{k}\rangle$. It
is usually called the computational basis. In the following, $\sigma_{\alpha}^{(k)}$
will denote a Pauli matrix ($\alpha=x,y,z$) acting on the $k$th
qubit.

A quantum computation consists in modifying the state $|\Psi\rangle$
by cleverly designing the time dependence of the Hamiltonian. In gate-based
quantum computers, the time evolution $U\equiv U(t)$ (Eq. (\ref{eq:U_dyson_series}),
we drop the $t$ dependence for clarity) is split into a sequence
\begin{equation}
U=\prod_{k=1}^{N_{g}}U_{k}\label{eq:gate_sequence}
\end{equation}
of $N_{g}$ elementary unitary operations $U_{k}$, referred to as
quantum ``gates'', that correspond to switching on some terms in
the Hamiltonian. For instance, in Fig. \ref{fig:Circuit-model.},
the $G$ gate, corresponding to a unitary $U_{G}$, may correspond
to a Hamiltonian $H_{G}(t)=\frac{\Omega(t)}{2}\sigma_{x}^{(1)}-\delta(t)\sigma_{z}^{(1)},$
with $U_{G}$ given by the Dyson series $U_{G}=Te^{-i\int_{0}^{t}H_{G}(\tau)d\tau}$.
Since $H_{G}$ acts only on the first qubit, so does $U_{G}$. Such
a single-qubit gate (depicted by a box acting only on one line, namely
one qubit) does not create entanglement (acting with $U_{G}$ on a
factorized state yields a factorized state). A standard single-qubit
gate is the Hadamard gate, whose expression in the $\left\{ |0\rangle,|1\rangle\right\} $
basis reads
\begin{equation}
U_{\mathrm{H}}=\frac{1}{\sqrt{2}}\left[\begin{array}{cc}
1 & 1\\
1 & -1
\end{array}\right].\label{eq:Hadamard}
\end{equation}
In other words, $U_{\mathrm{H}}|0\rangle=\frac{1}{\sqrt{2}}\left(|0\rangle+|1\rangle\right)$:
the Hadamard gate creates superpositions. Other important standard
single-qubit gates are single-qubit rotations, defined as
\begin{equation}
R_{\alpha}(\theta)=e^{-i\frac{\theta}{2}\sigma_{\alpha}}.\label{eq:rotation_gate}
\end{equation}
They rotate the qubit around the axis $\alpha$ by an angle $\theta$.

Gates depicted as boxes spanning several lines, on the other hand,
are many-qubit gates. They require Hamiltonians that couple several
qubits (like, say, $H(t)=g(t)\sigma_{x}^{(1)}\otimes\sigma_{x}^{(2)}$,
which generates a gate $U_{\mathrm{XX}}(t)=\exp\left(-i\int_{0}^{t}g(\tau)\mathrm{d}\tau\sigma_{x}^{(1)}\otimes\sigma_{x}^{(2)}\right)$).
The specific two-qubit gate shown in Fig. \ref{fig:Circuit-model.}
after gate $G$ is a so-called CNOT gate. Its expression in the $\left\{ |00\rangle,|01\rangle,|10\rangle,|11\rangle\right\} $
basis reads
\begin{equation}
U_{\mathrm{CNOT}}=\left[\begin{array}{cccc}
1 & 0 & 0 & 0\\
0 & 1 & 0 & 0\\
0 & 0 & 0 & 1\\
0 & 0 & 1 & 0
\end{array}\right].\label{eq:CNOT}
\end{equation}
For instance, $U_{\mathrm{CNOT}}|10\rangle=|11\rangle$: if the first
(``control'') qubit is in state $|1\rangle$, the second (``target'')
qubit is flipped. This notion of controlled gate can be generalized:
a controlled-$U$ gate $U_{c}$ is such that $U_{c}|0\rangle|\psi\rangle=|0\rangle|\psi\rangle$
and $U_{c}|1\rangle|\psi\rangle=|1\rangle U|\psi\rangle$.

Let us stress that several Hamiltonians $H(t)$ can yield the same
unitary evolution (for instance several functions $g(t)$ generate
the same $U_{\mathrm{XX}}$ gate in the example above, provided the
integral $\int_{0}^{t}g(\tau)\mathrm{d}\tau$ is the same). Thus,
going to a circuit-level description of the time evolution instead
of staying at the Hamiltonian level allows one to abstract away the
implementation details. Depending on the architecture, we may have
different $H(t)$ yielding the same unitary evolution. To benefit
from this description, we of course need to be able to switch on and
off terms of the Hamiltonian acting on selected qubits, a property
we called ``local control'' before, and that analog computers do
not possess. Finally, we call ``quantum circuit''\index{quantum circuits}
the graphical representation of the sequence of gates: it describes
the unitary operation that is effected by the quantum computer.

\subsubsection{Time evolution\label{subsec:Time-evolution:Trotter}}

In view of our goal of computing the Green's function (Eq. (\ref{eq:G_greater_explicit})),
there is a special unitary operation we would like to realize, namely
a time evolution (like $U=e^{-iH_{\mathrm{AIM}}t}$ in Eq. (\ref{eq:G_greater_explicit})).

Many quantum algorithms have been and are being developed to perform
this task, also known as the ``Hamiltonian simulation problem''.
With an analog quantum computer (or quantum simulator) that directly
implements the Hamiltonian under study (say $H_{\mathrm{hardware}}=H_{\mathrm{AIM}}$),
this just consists in letting the system evolve for a time $t$. For
gate-based quantum computers, we need to find a sequence of gates
that at least approximately implements $e^{-iHt}$ (the same method
applies to the more time-dependent case $Te^{-i\int_{0}^{t}H(\tau)d\tau}$).
The most straightforward algorithm, known as the Trotterization or
product formula method, relies on decomposing $H$ as a weighted sum
of (e.g ) products of Pauli matrices:
\begin{equation}
H=\sum_{i=1}^{M}\lambda_{i}P_{i},\label{eq:Pauli_decomp_H}
\end{equation}
with $\lambda_{i}\in\mathbb{R}$ and $P_{i}=\bigotimes_{k=1}^{N}\sigma_{\alpha_{k}}^{(k)}$,
where here the index $\alpha_{k}$ runs over $0,x,y,z$, where $\sigma_{0}=I$
by convention. Because the individual Pauli terms $P_{_{i}}$ do not
commute with one another, one then performs a similar time slicing
(called Trotterization\index{Trotterization}) as the one we encountered
in the TEBD algorithm in a tensor network context (see Eq. (\ref{eq:TEBD_trotter})):
\begin{equation}
e^{-iHt}=\left(e^{-iHt/N_{t}}\right)^{N_{t}}=\left(\prod_{i=1}^{M}e^{-i\frac{\lambda_{i}}{N_{t}}P_{i}t}+O\left(\frac{t}{N_{t}}\right)^{2}\right)^{N_{t}}=\prod_{m=1}^{N_{t}}\prod_{i=1}^{M}e^{-i\frac{\lambda_{i}}{N_{t}}P_{i}t}+O\left(\frac{t^{2}}{N_{t}}\right).\label{eq:trotter_time_ev}
\end{equation}
With this slicing, one obtains a sequence of $N_{t}\cdot M$ unitary
operators with the form of a Pauli rotation $R_{P_{i}}(\theta_{i})=e^{-i\frac{\theta_{i}}{2}P_{i}}$,
with $\theta_{i}=2\lambda_{i}\frac{t}{N_{t}}$. 

Let us know show that each $R_{P_{i}}(\theta_{i})$ is easy to implement
in terms of a simple quantum circuit.

\begin{figure}
\noindent \begin{centering}
\includegraphics[width=0.6\columnwidth]{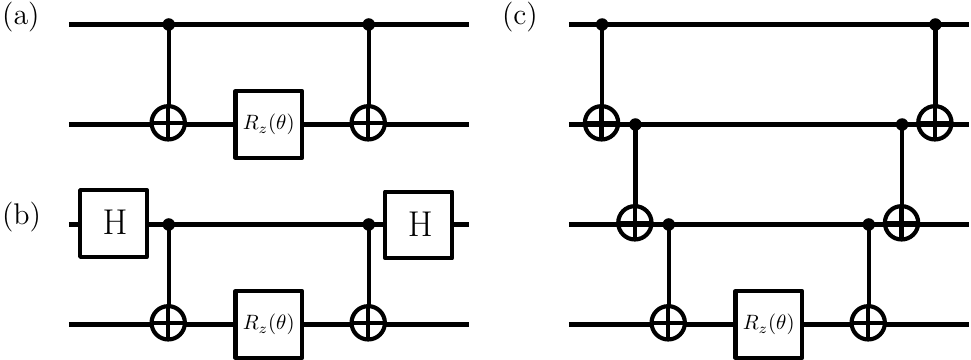}
\par\end{centering}
\caption{Circuits for $R_{P_{i}}(\theta)$. (a) Circuit for $R_{\sigma_{z}^{(1)}\sigma_{z}^{(2)}}(\theta)$.
(b) Circuit for $R_{\sigma_{x}^{(1)}\sigma_{z}^{(2)}}(\theta)$. (c)
Circuit for $R_{\sigma_{z}^{(1)}\sigma_{z}^{(2)}\sigma_{z}^{(3)}\sigma_{z}^{(4)}}(\theta)$.
\label{fig:Circuit-for-RZZ}}
\end{figure}
Let us start with a simple case $P_{i}=\sigma_{z}^{(1)}\sigma_{z}^{(2)}$.
It is easy to check that the circuit shown in Fig. \ref{fig:Circuit-for-RZZ}
implements $R_{\sigma_{z}^{(1)}\sigma_{z}^{(2)}}(\theta)$. One easily
shows, by induction, that $R_{\bigotimes_{m=1}^{K}\sigma_{z}^{(k_{m})}}(\theta)$
is implemented by a circuit with CNOTS on pairs $(k_{1},k_{2})$,
$(k_{2},k_{3})$, ..., $(k_{K-1},k_{K})$, a rotation $R_{z}(\theta)$
on the $K$th qubit, and then the reversed sequence of CNOTs, as illustrated
in Fig. \ref{fig:Circuit-for-RZZ} (c). Finally, to perform a general
Pauli rotation (with Paulis on the $x$ and $y$ axis), one performs
single-qubit rotations on the requisite qubits to go back to the $z$
axis. For instance, for $P_{i}=\sigma_{x}^{(1)}\sigma_{z}^{(2)}$,
noticing that
\begin{equation}
U_{\mathrm{H}}\sigma_{z}U_{\mathrm{H}}^{(\dagger)}=\sigma_{x},\label{eq:z_to_x_identity}
\end{equation}
we see that
\begin{align*}
U_{\mathrm{H}}^{(1)}R_{\sigma_{z}^{(1)}\sigma_{z}^{(2)}}(\theta)U_{\mathrm{H}}^{(1)} & =U_{\mathrm{H}}^{(1)}\left(\cos(\theta)I-i\sin(\theta)\sigma_{z}^{(1)}\sigma_{z}^{(2)}\right)U_{\mathrm{H}}^{(1)}\\
 & =\cos(\theta)I-i\sin(\theta)\sigma_{x}^{(1)}\sigma_{z}^{(2)}\\
 & =R_{\sigma_{x}^{(1)}\sigma_{z}^{(2)}}(\theta).
\end{align*}
 This is illustrated in Fig. \ref{fig:Circuit-for-RZZ} (b).

We thus have a generic, approximate way to perform the time evolution
of a Hamiltonian that we have decomposed as a sum of Pauli operators
(Eq. (\ref{eq:Pauli_decomp_H})). Let us now count the number of operations
contained in this circuit: there are $N_{t}$ trotter steps, each
of which contains $M$ operators of the form $R_{P_{i}}(\theta)$.
The number of gates in each operator depends on the support $s_{i}$
(number of non-identity Pauli operators) in $P_{i}$: we have $O(s_{i})$
gates in the corresponding subcircuit. If we call $s=\text{\ensuremath{\max_{i}s_{i}}}$,
we obtain $N_{g}=O(N_{t}Ms)$, with a total error $\epsilon=O(t^{2}/N_{t})$.
One usually wants to adjust $N_{t}$ to reach a desired error $\epsilon$:
$N_{t}=O(t^{2}/\epsilon)$ and thus, 
\begin{equation}
N_{g}=O(Mt^{2}s/\epsilon).\label{eq:Ng_Trotter}
\end{equation}
This calls forth several remarks: (1) the number of terms $M$ needs
to be ``small''. A generic Hamiltonian in $\mathcal{H}_{N}$ will
have $M=2^{N}$, which is intractable. However, as we shall see, the
Hamiltonians we are interested in have a number of terms $M$ that
scales polynomially with $N$ (which is considered to be tractable);
(2) the longer the evolution time $t$, the more gates. Here, the
quadratic dependence (instead of the expected linear dependence) comes
from the fact we had to slice the time evolution. We can use higher-order
Suzuki-Trotter formula that will lead to a scaling that is closer
to linear, but with an overhead in the number of gates per slice (see
\cite{Childs2019} for a general upper bound on the Trotter error).
More advanced methods such as qubitization (\cite{HaoLow2017}) achieve
a linear scaling , but they usually require additional (``ancilla'')
qubits. They can also achieve a better scaling than $1/\epsilon$;
(3) finally, the support $s$ of the terms in the decomposition of
$H$ plays an important role. As we shall see below, the fermionic
nature of the Hamiltonian of interest here leads to quite large supports.

\subsubsection{Quantum measurements and the variance issue\label{subsec:Quantum-measurements-and}}

Before we turn to the specifics of fermionic problems on qubit quantum
computers, let us examine one last issue with the quantity we need
to compute, Eq. (\ref{eq:G_greater_explicit}). It is expressed as
a quantum mechanical average value $\langle\psi|\hat{O}|\psi\rangle$,
with operator $\hat{O}=U^{\dagger}c_{\sigma}Uc_{\sigma}^{\dagger}$
and $U=e^{-iH_{\mathrm{AIM}}t}$. We have seen in the previous section
how to perform $U$ as a quantum circuit, but what quantum mechanics
gives us easy access to is the estimation of $\langle\psi|\hat{A}|\psi\rangle$,
where $\hat{A}$ is a Hermitian operator, not any operator like $\hat{O}$.

Before we turn to specific quantum circuits that allow us to get access
to $\langle\psi|\hat{O}|\psi\rangle$, let us focus on the Hermitian
case. In quantum mechanics, the measurement of an observable $\hat{A}$
with eigenvalues $\lambda$ and eigenvectors $|\varphi_{\lambda}\rangle$
results (in the nondegenerate case) in a collapse of $|\psi\rangle$
to one of the eigenvectors $|\varphi_{\lambda}\rangle$ with a probability
given by Born's rule, $p(\lambda)=\left|\langle\varphi_{\lambda}|\psi\rangle\right|^{2}$,
and gives access to the value $\lambda$ of the corresponding eigenvalue.
Repeating the circuit and measurement several (say $N_{s}$) times
(usually called ``shots'') allows to compute a statistical estimator
$\overline{A}(N_{s})$ that converges to the expectation value of
$\hat{A}$ in state $|\psi\rangle$ in the large $N_{s}$ limit:
\begin{equation}
\overline{A}(N_{s})=\frac{1}{N_{s}}\sum_{i=1}^{N_{s}}\lambda_{i}\xrightarrow[N_{s}\infty]{}\sum_{\lambda}p(\lambda)\lambda=\langle\psi|\hat{A}|\psi\rangle=\langle A\rangle\label{eq:avg_value}
\end{equation}

Quantum computers can either be used in a one-shot way or to compute
averages. The one-shot way is particularly useful when the distribution
$p(\lambda)$ is peaked around a value $\lambda_{0}$ which one wants
to discover: in that case, a single (or a few) shots will yield $\lambda_{0}$.
As for the computation of averages, it also works when $p(\lambda)$
is not peaked, but one has to contend with the statistical error (standard
error on the mean)
\begin{equation}
\Delta A(N_{s})=\sqrt{\langle\left(\langle A\rangle-\overline{A}(N_{s})\right)^{2}\rangle}=\sqrt{\frac{\mathrm{Var}(A)}{N_{s}}},\label{eq:standard_error_on_mean}
\end{equation}
where the second equality holds because independent experiments (shots)
are iid. The variance is, explicitly,
\[
\mathrm{Var}(A)=\langle\psi|A^{2}|\psi\rangle-\left(\langle\psi|A|\psi\rangle\right)^{2},
\]
so the number of shots needed to attain a fixed accuracy $\Delta A$
depends both on the observable $A$ to be measured and the state $|\psi\rangle$
on which it is measured. For instance, if one is trying to generate
the ground state $|\Psi_{0}\rangle$ of some Hamiltonian $H$, and
wants to measure its energy $\langle H\rangle$, then, because $|\Psi_{0}\rangle$
is an eigenstate of $H$, the variance will vanish and only one shot
will suffice. This property, called variance reduction, is often exploited
in (classical) variational Monte-Carlo algorithms: as the algorithm
converges to the ground state, a fixed number of samples (shots) gives
an increased accuracy.

\begin{figure}
\noindent \begin{centering}
\includegraphics[width=0.7\columnwidth]{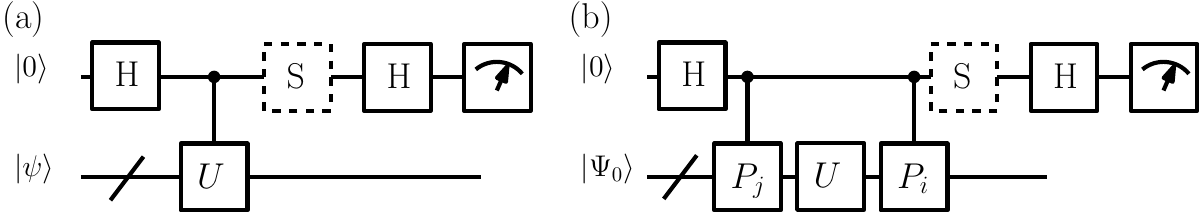}
\par\end{centering}
\caption{Hadamard test: (a) Computation of $\langle\psi|U|\psi\rangle$ by
the Hadamard test circuit. The $S$ gate is used only to compute the
imaginary part. (b) Application to the computation of $\langle\Psi_{0}|U^{\dagger}P_{i}UP_{j}|\Psi_{0}\rangle$
terms. \label{fig:Hadamard_test}}
\end{figure}

However, not every observable $\hat{A}$ can be measured on quantum
computers. In fact, in most technologies, only Pauli-z matrices $\sigma_{z}^{(k)}$,
for $k=1\dots N$, and tensor products thereof, can be measured. This
is represented by a meter symbol in quantum circuits (see the symbols
at the end of the first lines of Fig. \ref{fig:Hadamard_test}(a)
and (b)). Other Pauli expectation values can be computed by rotating
the state to the right axis: for instance, to measure $\langle\sigma_{x}\rangle=\langle\psi|\sigma_{x}|\psi\rangle$,
one can apply a Hadamard gate to get $|\psi'\rangle=U_{\mathrm{H}}|\psi\rangle$
and then, using (\ref{eq:z_to_x_identity}), measure $\langle\sigma_{x}\rangle'=\langle\psi'|\sigma_{x}|\psi'\rangle$.

As for generic, multi-qubit observables $\hat{A}$, they can, as we
saw earlier for the Hamiltonian, always be decomposed as a weighted
sum of products of Pauli matrices:
\begin{equation}
\hat{A}=\sum_{i=1}^{M}\lambda_{i}P_{i}.\label{eq:Pauli_decomp}
\end{equation}
One can thus in principle compute $\langle A\rangle$ by computing
separately the $M$ terms $\langle P_{i}\rangle$. For example, to
measure $\langle\sigma_{x}^{(1)}\sigma_{z}^{(2)}\rangle$, one first
applies a Hadamard gate on the first qubit, and then measures the
operator $\sigma_{z}^{(1)}\otimes\sigma_{z}^{(2)}$ repeatedly.

We now turn to the problem at hand, namely computing $\langle\psi|\hat{O}|\psi\rangle$
for a generic operator. A generic strategy is to decompose $\hat{O}$
as a sum of unitaries, $\hat{O}=\sum_{i}\lambda_{i}\hat{U}_{i}$ (with
$\lambda_{i}\in\mathbb{C}$), and compute $\langle\psi|\hat{U}_{i}|\psi\rangle$
separately. (Note that a Pauli decomposition like Eq. (\ref{eq:Pauli_decomp})
is also a decomposition as a linear combination of unitaries since
Pauli matrices are unitary).

The circuit showed in Fig. \ref{fig:Hadamard_test}(a), dubbed a Hadamard
test\index{Hadamard test}, does the job of computing $\langle\psi|\hat{U}|\psi\rangle$
for a given unitary $U$: starting from a state $|0\rangle\otimes|\psi\rangle$,
applying the first Hadamard gate yields $\frac{|0\rangle+|1\rangle}{\sqrt{2}}\otimes|\psi\rangle$.
Then, the controlled unitary yields $\frac{1}{\sqrt{2}}\left(|0\rangle|\psi\rangle+|1\rangle U|\psi\rangle\right)$,
and the final Hadamard yields the final state $|\psi_{f}\rangle=\frac{1}{2}\left(|0\rangle\left(I+U\right)|\psi\rangle+|1\rangle\left(I-U\right)|\psi\rangle\right)$.
The meter symbol stands for a $\sigma_{z}^{(0)}$ measurement. The
probability of getting outcome $0$ is given by Born's rule $p(0)=\langle\psi_{f}|\left(|0\rangle\langle0|\otimes I\right)|\psi_{f}\rangle=\frac{1}{4}\left\Vert \left(I+U\right)|\psi\rangle\right\Vert ^{2}=\frac{1}{2}\left(1+\mathrm{Re}\langle\psi|U|\psi\rangle\right)$,
so that 
\begin{equation}
\mathrm{Re}\langle\psi|U|\psi\rangle=1-2p(0).\label{eq:U_avg_hadamard}
\end{equation}
If one adds a $S$ gate to the circuit (dashed box in Fig. \ref{fig:Hadamard_test}(a),
whose matrix is $U_{S}=\left[\begin{array}{cc}
1 & 0\\
0 & i
\end{array}\right]$), $U$ changes to $iU$ in the above computation, giving $\mathrm{Im}\langle\psi|U|\psi\rangle=2p(0)-1$.
Thus, provided the ability to implement a controlled-$U$ evolution,
we can compute terms of the form $\langle\psi|U|\psi\rangle$, and
thus arbitrary values $\langle\psi|\hat{O}|\psi\rangle$ via the linear
combination of unitaries.

The last step we have to go through is how to express $\hat{O}=U^{\dagger}c_{\sigma}Uc_{\sigma}^{\dagger}$
as a linear combination of unitaries. For this, we need to go from
fermionic operators to qubit ones.

\subsubsection{Quantum computing for fermions\label{subsec:QC-for-fermions}}

In the previous sections, we have learned to perform time evolutions
of Hamiltonian operators decomposed as a weighted sum of Pauli operators
(Eq. (\ref{eq:Pauli_decomp_H})), and to measure quantities of the
form $\langle\psi|\hat{O}|\psi\rangle$ given a decomposition of $\hat{O}$
as a weighted sum of unitaries. We now explain how to obtain the precise
form of these decompositions in the case where $H$ is a fermionic
operator like the AIM, and $\hat{O}=U^{\dagger}c_{\sigma}Uc_{\sigma}^{\dagger}$.

The task consists in finding a mapping between a Fock space with $N$
fermionic orbitals, with a Fock basis $\left\{ |n_{1},\dots,n_{N}\rangle=\prod_{k=1}^{N}\left(c_{k}^{\dagger}\right)^{n_{k}}|0,\dots,0\rangle,n_{k}\in\left\{ 0,1\right\} \right\} $,
and a Hilbert space with $N$ qubits, with a basis $\left\{ |b_{1}\rangle\otimes|b_{2}\rangle\otimes\cdots\otimes|b_{N}\rangle,b_{k}\in\left\{ 0,1\right\} \right\} $.

\paragraph{The Jordan-Wigner mapping}

The simplest mapping (also called encoding) between these two bases
consists in defining, for all $k=1\dots N$,
\begin{equation}
b_{k}=n_{k}.\label{eq:jw_encoding}
\end{equation}
It is called the Jordan-Wigner encoding\index{Jordan-Wigner encoding}.
Having chosen this correspondence between states, we want to find
the qubit operator $\tilde{c}^{\dagger}$ that acts on the $|b_{1},\dots,b_{N}\rangle$
the same way as $c^{\dagger}$ acts on the $|n_{1},\dots,n_{N}\rangle$,
namely:
\begin{align}
c_{k}^{\dagger}|n_{1},\dots,0,\dots,n_{N}\rangle & =\left(-\right)^{\sum_{k'=1}^{k-1}n_{k'}}|n_{1},\dots,1,\dots,n_{N}\rangle\label{eq:creation_op_action}\\
c_{k}^{\dagger}|n_{1},\dots,1,\dots,n_{N}\rangle & =0\nonumber 
\end{align}
The operator that turns a two-level system $|0\rangle$ into $|1\rangle$
and vanishes on $|1\rangle$ is $\sigma_{+}=\left(\begin{array}{cc}
0 & 0\\
1 & 0
\end{array}\right)=\frac{1}{2}\left(\sigma_{x}-i\sigma_{y}\right)$. To take into account the $\left(-\right)^{\sum_{k'=1}^{k-1}n_{k'}}$
factor, one needs to introduce $\sigma_{z}$ operators on qubits $k'<k$,
so that:
\begin{equation}
\tilde{c}_{k}^{\dagger}=\sigma_{z}^{(1)}\cdots\sigma_{z}^{(k-1)}\sigma_{+}^{(k)}.\label{eq:jw_def}
\end{equation}
For a spinful fermionic model, we further need to pick an ordering
of the $(i,\sigma)$ index. Two natural choices arise: either $(1,\uparrow),(1,\downarrow),(2,\uparrow),(2,\downarrow),\dots$
or $(1,\uparrow),(2,\uparrow),\dots,(1,\downarrow),(2,\downarrow),\dots$.
These choices will yield equivalent, but different quantum circuits.
Choosing the second ordering (the one that does not mix different
spins), a $c_{i\sigma}^{\dagger}c_{j\sigma}$ term appearing in the
kinetic term of the Hubbard model will transform into
\begin{equation}
\tilde{c}_{i\sigma}^{\dagger}\tilde{c}_{j\sigma}=\sigma_{+}^{(i\sigma)}\sigma_{z}^{(i+1,\sigma)}\cdots\sigma_{z}^{(j-1,\sigma)}\sigma_{-}^{(j\sigma)}\label{eq:kinetic_term_jw}
\end{equation}
while Hubbard interaction terms like $n_{i\uparrow}n_{i\downarrow}$
will transform into $\tilde{n}_{i\uparrow}\tilde{n}_{i\downarrow}=\left(\frac{I-\sigma_{z}^{(i\uparrow)}}{2}\right)\left(\frac{I-\sigma_{z}^{(i\downarrow)}}{2}\right).$

In other words, the Anderson Hamiltonian, with terms acting on at
most 2 fermionic orbitals at the same time, is turned into a spin
Hamiltonian with terms acting on up to $2N_{b}$ orbitals due to the
strings of $\sigma_{z}$ operators, called Jordan-Wigner strings,
appearing in (\ref{eq:kinetic_term_jw}). Explicitly: mapping the
impurity orbital ($c_{\sigma}^{\dagger}$) to spin index $1$ and
the $k$th bath orbital ($a_{k\sigma}^{\dagger}$) to spin index $1+k$
we obtain:

\begin{align}
\tilde{H}_{\mathrm{AIM}} & =\left(\frac{U}{4}-\frac{\mu}{2}+\sum_{k\sigma}\frac{\epsilon_{k}}{2}\right)I-\sum_{k\sigma}\frac{\epsilon_{k}}{2}\sigma_{z}^{(1+k,\sigma)}+\left(\frac{\mu}{2}-\frac{U}{4}\right)\sum_{\sigma}\sigma_{z}^{(1,\sigma)}+\frac{U}{4}\sigma_{z}^{(1,\uparrow)}\sigma_{z}^{(1,\downarrow)}\nonumber \\
 & \;\;+\sum_{k\sigma}\frac{V_{k}}{4}\left(\left(\sigma_{x}^{(1\sigma)}-i\sigma_{y}^{(1\sigma)}\right)\sigma_{z}^{(2,\sigma)}\cdots\sigma_{z}^{(k,\sigma)}\left(\sigma_{x}^{(1+k,\sigma)}+i\sigma_{y}^{(1+k,\sigma)}\right)+\mathrm{h.c}\right)\label{eq:H_AIM_spin}
\end{align}
We now have a Pauli decomposition of $\tilde{H}_{\mathrm{AIM}}$ with
a total of $2(N_{b}+1)$ 1-qubit terms (disregarding the identity
term), one two-qubit term stemming from the Hubbard interaction, and
$8N_{b}$ terms stemming from the hybridization term, with supports
ranging from 2 to $N_{b}+1$.

Using this decomposition, we can apply the Trotterization method introduced
above to construct a quantum circuit that approximates $e^{-iH_{\mathrm{AIM}}t}$.
One Trotter step will consist in applying $R_{z}(\theta)$ rotation
gates to all qubits (with angles given by the prefactors in the terms
of the first line of expression (\ref{eq:H_AIM_spin})). As for the
second line of (\ref{eq:H_AIM_spin}), it gives rise to four subcircuits
of the type $R_{P_{i}}(\theta)$ that we described in subsection \ref{subsec:Time-evolution:Trotter},
with circuits of length up to $O(N_{b}),$and thus a general scaling
of $N_{g}=O(Mt^{2}s/\epsilon)=O\left(N_{\mathrm{b}}^{2}t^{2}/\epsilon\right)$
for the time evolution circuit.

Expression (\ref{eq:jw_def}) also allows us to express $\hat{O}_{\sigma}=U^{\dagger}c_{\sigma}Uc_{\sigma}^{\dagger}$
as a linear combination of unitaries: replacing the creation and annihilation
operators by (\ref{eq:jw_def}), we obtain:
\begin{align}
\hat{O}_{\uparrow} & =\frac{1}{4}U^{\dagger}\left(\sigma_{x}^{(1,\uparrow)}+i\sigma_{y}^{(1,\uparrow)}\right)U\left(\sigma_{x}^{(1,\uparrow)}-i\sigma_{y}^{(1,\uparrow)}\right)\nonumber \\
 & =\frac{1}{4}\left(U^{\dagger}\sigma_{x}^{(1,\uparrow)}U\sigma_{x}^{(1,\uparrow)}-iU^{\dagger}\sigma_{x}^{(1,\uparrow)}U\sigma_{y}^{(1,\uparrow)}+iU^{\dagger}\sigma_{y}^{(1,\uparrow)}U\sigma_{x}^{(1,\uparrow)}+U^{\dagger}\sigma_{y}^{(1,\uparrow)}U\sigma_{y}^{(1,\uparrow)}\right),\label{eq:O_as_LCU}
\end{align}
which is in the requisite form.

We thus have achieved our goal. We note that in the single-impurity
case that we have picked as an example, we could have ``linearized''
the bath, namely transformed the current problem, where the impurity
is hybridized to every bath site (a so-called ``star geometry''),
to a problem where the $\uparrow$ impurity site is hybridized only
to one bath site (itself hybridized to a single other bath site and
so on and so forth), and likewise for the $\downarrow$ impurity site.
This would have allowed for a ``chain geometry'' where the Jordan-Wigner
strings disappear, yielding a scaling of $N_{\mathrm{g}}=O\left(N_{\mathrm{b}}t^{2}/\epsilon\right)$
gates for the time evolution circuit. However, in the $N_{c}>1$ case,
we can no longer perform this trick.

\paragraph{Other encodings}

The Jordan-Wigner encoding is not the only possible fermion-to-spin
mapping. While the Jordan-Wigner encoding stores all the information
about the orbital occupation in the states (see Eq. (\ref{eq:jw_encoding}))
and the information about the parity in the operators (see the string
of $\sigma_{z}$ operators in (\ref{eq:jw_def})), a mirror encoding
called the parity encoding does the reverse. For instance, in the
parity encoding, $b_{k}=\sum_{k'=1}^{k}n_{k}\;\mathrm{mod.\,2}$ namely
the states store the parity information, and likewise a string of
Pauli operators in the expression for the operators will store the
occupation... yielding, as for the Jordan-Wigner transformation, terms
with a support $O(N)$ (with $N$ the number of fermionic orbitals)
in the Pauli decomposition of the corresponding qubit Hamiltonian.
A more sophisticated encoding called the Bravyi-Kitaev encoding (\cite{Bravyi2002,Havlicek2017})
mixes the occupation and parity information in a tree structure that
allows one to obtain terms with $O(\log(N))$ support. Let us also
note that while the aforementioned encodings use a Hilbert space of
the same size as the fermionic Fock space, other encodings use larger
Hilbert spaces (and thus more qubits than fermionic modes) to obtain
local qubit Hamiltonians (at the expense of ancillary qubits) (\cite{Verstraete2005,Derby2021,Algaba2023}).

\subsection{Quantum algorithms for Green's functions}

In the previous section, we introduced the main quantum computing
building blocks to deal with fermionic models. In this section, we
use these building blocks to introduce several methods to compute
Green's functions with gate-based quantum computers.

\subsubsection{Ground state preparation: the example of the adiabatic method\label{subsec:Ground-state-preparation:adiab}}

The greater Green's function $G^{>}(t)$ (see Eq. (\ref{eq:G_greater_explicit})
for a definition) requires the preparation of the ground state $|\Psi_{0}\rangle$
(at zero temperature) or the Gibbs state $\rho=e^{-\beta H}/Z$ at
finite temperature. In this section we tackle only the zero-temperature
case, although quantum algorithms have also been developed for Gibbs
state preparation, usually with the help of additional qubits.

\begin{figure}
\noindent \begin{centering}
\includegraphics[width=0.8\columnwidth]{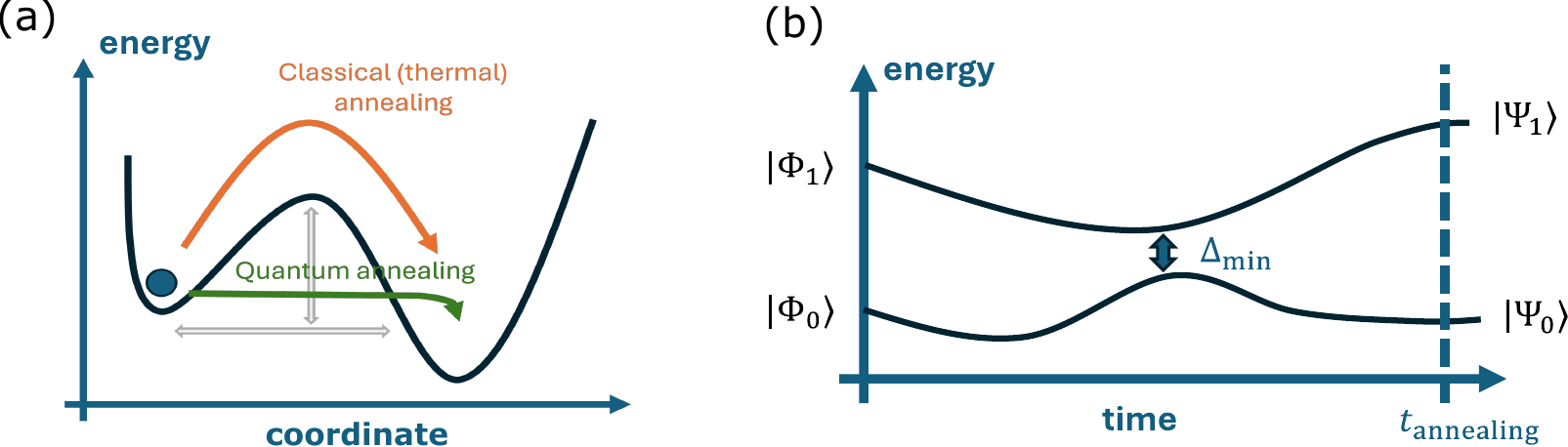}
\par\end{centering}
\caption{Adiabatic quantum annealing method. (a) Quantum tunneling (green)
vs thermal hopping (orange). (b) Time evolution of the instantaneous
energy levels and definition of the minimum gap.\label{fig:Adiabatic-quantum-annealing}}
\end{figure}

A standard method to perform this state preparation is the adiabatic
method\index{adiabatic method}. It consists initializing the qubit
register in an easy-to-prepare state $|\Phi_{0}\rangle$ that should
also be the ground state of a Hamiltonian $H_{0}$. This Hamiltonian,
sometimes called the mixer Hamiltonian, must be simple enough that
its ground state $|\Phi_{0}\rangle$ is easy to prepare, and it must
be such that it couples the different (usually unknown) eigenstates
of the Hamiltonian $H$ whose ground state $|\Psi_{0}\rangle$ we
want to prepare. One then slowly deforms $H_{0}$ into $H$ over a
period of time $t_{\mathrm{annealing}}$, for instance
\begin{equation}
H(t)=\left(1-t/t_{\mathrm{annealing}}\right)H_{0}+t/t_{\mathrm{annealing}}H.\label{eq:annealing_schedule}
\end{equation}
In this time evolution, the role of $H_{0}$ can be seen as allowing
for tunneling events between local minimal of the energy landscape
of $H$, as pictorially represented in Fig. \ref{fig:Adiabatic-quantum-annealing}
(a). Over time, $H_{0}$ is slowly turned off, hence the name ``quantum
annealing'' given to this method, in analogy to the classical (thermal)
annealing method, which consists in slowly lowering the temperature.
The adiabatic theorem (see \cite{Albash2018}) guarantees that for
long enough annealing times the system remains in the instantaneous
ground state of $H(t)$. More precisely, 
\begin{equation}
t_{\mathrm{annealing}}\gg\frac{V}{\Delta_{\mathrm{min}}^{2}},\label{eq:adiabatic_condition}
\end{equation}
with $V$ a matrix element, and $\Delta_{\mathrm{min}}$ the minimum
gap over time between the instantaneous ground state of $H(t)$ and
the first excited state, as represented in Fig. \ref{fig:Adiabatic-quantum-annealing}
(b). 

Thus, to prepare the ground state $|\Psi_{0}\rangle$ using the quantum
annealing method, we need to construct a circuit that implements the
time evolution $H(t)$ for a long enough $t_{\mathrm{annealing}}$.
The construction of the quantum circuit to perform this time evolution
uses the methods described in subsection \ref{subsec:Time-evolution:Trotter},
for instance Trotterization. We note that a longer $t_{\mathrm{annealing}}$
will generically lead to a deeper state preparation circuit: the best
Hamiltonian evolution methods (like qubitization) achieve a linear
scaling of the number of gates as a function of the evolution time.
This scaling is optimal owing to the so-called no-fast-forwaring theorem,
which stipulates that time evolutions of sparse Hamiltonians of duration
$t$ will require a number of gates at best linear in $t$ \cite{Berry2006}.
As a consequence, an important practical question is the size $\Delta_{\mathrm{min}}$,
which itself depends on $H_{0}$, $H$ and the shape of the interpolation
between both. It is in general hard to compute, but we at least know
that $\Delta_{\mathrm{min}}\geq\Delta$, with $\Delta$ the gap of
$H$. Therefore, trying to prepare the ground state of a small-gap
Hamiltonian (and for that matter a gapless one, like a metal) will
take very long circuits.

There exists heuristic alternatives around adiabatic quantum annealing
that set out to find shortcuts to adiabaticity (see \cite{Guery-Odelin2019}
for a review). Variational methods like the ones we will introduce
below (see subsection \ref{subsec:Near-term-algorithms.-Variationa})
can be regarded as examples of such heuristics.

\paragraph{The complexity of ground state preparation}

On a more formal level, let us note that the generation of the ground
state of a many-body Hamiltonian is generally expected to be hard
(that is, exponential) even on quantum computers: it has been proven
to be ``QMA-complete'' for $k\geq2$-local Hamiltonians \cite{Kempe2008}
(and hence for the Hubbard model \cite{Schuch2009}). Here, $k$ denotes
the maximal support of the terms appearing in e.g the Pauli decomposition
of $H$, and QMA---for quantum Merlin-Arthur---is a quantum analog
of the NP class. Impurity problems may be less difficult in terms
of ground state preparation: they are in the QCMA (for quantum-classical
Merlin-Arthur) class \cite{Bravyi2016a}, a class that stands between
NP and QMA in terms of hardness. If they are gapped, they become easy
(efficiently solvable) both for classical and quantum computers \cite{Bravyi2016a}.
If not, recent work claims evidence that their ground state can be
efficiently prepared by a quantum computer, while remaining hard for
classical computers \cite{Erakovic2025}. This would provide an a
posteriori justification of quantum embedding methods like DMFT: using
locality properties of the interaction, they would help reduce an
exponentially problem (the Hubbard model) to an efficiently solvable
problem (the impurity problem).

\subsubsection{Time domain: Hadamard test \label{subsec:Time-domain:-Hadamard}}

Once the ground state is prepared, the Hadamard test method to compute
the greater Green's function $G^{>}(t)$ (Eq. (\ref{eq:G_greater_explicit}))
is a straightforward application (first introduced by \cite{Ortiz2001},
and well summarized in \cite{Wecker2015a}) of the few methods we
just introduced.

Using the decomposition (\ref{eq:O_as_LCU}) of $U^{\dagger}c_{\sigma}Uc_{\sigma}^{\dagger}$,
we see that computing $G^{>}(t)$ amounts to summing four contributions
of the form $\langle\Psi_{0}|U^{\dagger}P_{i}UP_{j}|\Psi_{0}\rangle$,
with $P_{i,j}\in\left\{ \sigma_{x}^{(1\uparrow)},\sigma_{y}^{(1\uparrow)}\right\} $.
Since $U^{\dagger}P_{i}UP_{j}$ is itself unitary, we only have to
use the Hadamard test depicted in Fig. \ref{fig:Hadamard_test}, with
a controlled-$U^{\dagger}P_{i}UP_{j}$ operation. A naive implementation
of this operation would be a sequence of a controlled $P_{j}$, controlled
$U$, controlled $P_{i}$ and then controlled $U^{\dagger}$. Yet,
the precise structure of the $U^{\dagger}P_{i}U$ allows for one simplification:
the controls on the time evolution $U$ and its inverse $U^{\dagger}$
can be removed. Let us show that we get the same outcome with and
without controls: with controls, if the ancilla qubit is in state
$|0\rangle$ and the state register in state $|\psi\rangle$, we end
up with $|0\rangle|\psi\rangle$. If the ancilla is in state $|1\rangle$,
we end up with $|1\rangle U^{\dagger}P_{i}U|\psi\rangle$. Without
controls, if the ancilla qubit is in state $|0\rangle$ and the state
register in state $|\psi\rangle$, we end up with $|0\rangle U^{\dagger}U|\psi\rangle=|0\rangle|\psi\rangle$.
If the ancilla is in state $|1\rangle$, we end up with $|1\rangle U^{\dagger}P_{i}U|\psi\rangle$.
QED.

The corresponding circuit can be further simplified by noticing that
the $U^{\dagger}$ operation has no influence on the final measurement
as it commutes with the remaining operations (that act only on the
ancilla qubit). It can thus be removed, which yields the circuit shown
in Fig. \ref{fig:Hadamard_test} (b): this circuit makes it clear
that computing the Green's function amounts to first generating the
ground state $|\Psi_{0}\rangle$, coupling it to an ancilla qubit
(controlled-$P_{j}$ operation), time evolving the system (operation
$U$), and then coupling again to the ancilla (controlled-$P_{i}$
operation) and measuring the ancilla.

Let us end this subsection by estimating the scaling of the run time
of the Hadamard test method with respect to the precision $\epsilon$
achieved on the Green's function. Since it requires an estimate of
$p(0)$ (Eq. (\ref{eq:U_avg_hadamard})), it is characterized by a
$1/\epsilon^{2}$ scaling: the statistical error on $p(0)$ scales
as $\epsilon\sim1/\sqrt{N_{\mathrm{samples}}}$, with $N_{\mathrm{samples}}$
the number of repetitions of the circuit to estimate the probability
of getting a $0$ outcome. Therefore, the run time scales as $N_{\mathrm{samples}}\sim1/\epsilon^{2}$.
This is typical of classical Monte-Carlo methods, and is a direct
consequence of the central limit theorem. As we shall see in the next
subsection, this scaling can be improved by resorting to quantum interferences.

\subsubsection{Frequency domain: quantum phase estimation\label{subsec:Frequency-domain:-quantum}}

\paragraph{High-level description}

Another quantum algorithm allows one to compute the Green's function
directly in the frequency domain, and with a better accuracy scaling.
Starting from Eq. (\ref{eq:G_greater_2}), and using $\int_{0}^{\infty}\mathrm{d}te^{i\omega t}=\pi\delta(\omega)+i\mathcal{P}\frac{1}{\omega}$
($\mathcal{P}$ denotes the Cauchy principal value), we obtain
\begin{align}
G^{>}(\omega) & =\sum_{\alpha}\left(-i\pi\delta\left(\omega-\left(E_{\alpha}-E_{0}\right)\right)+\mathcal{P}\left(\frac{1}{\omega-\left(E_{\alpha}-E_{0}\right)}\right)\right)\left|\langle\Psi_{\alpha}|c_{\sigma}^{\dagger}|\Psi_{0}\rangle\right|^{2}.\label{eq:G_greater_spectral}
\end{align}
The imaginary part of $G^{>}(\omega)$ is therefore a distribution
of peaks of weight $\left|\langle\Psi_{\alpha}|c_{\sigma}^{\dagger}|\Psi_{0}\rangle\right|^{2}$
at locations $E_{\alpha}-E_{0}$.

One of the most important quantum algorithms, quantum phase estimation
(QPE, \cite{Kitaev1995}), can be applied to this problem, as first
pointed out by \cite{Wecker2015a}. Given a unitary operator $U=e^{iHt}$
and one of its eigenvectors $|\Psi_{\alpha}\rangle$ with eigenvalue
$e^{iE_{\alpha}t}$, it returns, in a single repetition of the QPE
circuit, a $m$-bit estimate $\hat{E}_{\alpha}$ of $E_{\alpha}$
($t$ is adjusted so that $E_{\alpha}t$ is normalized: resolving
smaller energies requires longer times). When starting from any state
$|\Psi\rangle$ as an input, it returns a $m$-bit estimate $\hat{E}_{\alpha}$
of one of the eigenvalues $E_{\alpha}$ and projects the input state
to $|\Psi_{\alpha}\rangle$ with probability $\left|\langle\Psi_{\alpha}|\Psi\rangle\right|^{2}$.
Therefore, several calls to QPE on an input state $|\Psi\rangle$
will return a histogram with peaks of height $\left|\langle\Psi_{\alpha}|\Psi\rangle\right|^{2}$
at location $\hat{E}_{\alpha}$. We thus see that applying QPE to
the input state $c_{\sigma}^{\dagger}|\Psi_{0}\rangle$ and with $H=H_{\mathrm{AIM}}-E_{0}$
will directly yield an estimate of $\mathrm{Im}G^{>}(\omega)$ (and
hence, via Kramers-Kronig relations, of its real part).

In practice, however, $c_{\sigma}^{\dagger}|\Psi_{0}\rangle$ is not
a properly normalized state because $c_{\sigma}^{\dagger}$ is not
unitary. However, in a Jordan-Wigner encoding, assuming the impurity
orbital is the first orbital in the orbital order, $c_{\sigma}^{\dagger}+c_{\sigma}$
maps to $\tilde{c}_{\sigma}^{\dagger}+\tilde{c}_{\sigma}=\sigma_{x}^{(1)}$,
which is a bona fide unitary gate. Since the QPE circuit contains
gates that conserve the electron number, an input state $c_{\sigma}^{\dagger}|\Psi_{0}\rangle+c_{\sigma}|\Psi_{0}\rangle$
will be projected onto eigenstates $|\Psi_{\alpha}\rangle$ of $H$
with $N_{e}+1$ or $N_{e}-1$ electrons (where $N_{e}$ is the number
of electrons in the ground state $|\Psi_{0}\rangle$). To obtain the
histogram of the greater Green's function, we need to take into account
only those measurements that stem from $N_{e}+1$ states. To select
them, we just need to measure the number of electrons in the final
state.

\begin{figure}
\noindent \begin{centering}
\includegraphics[viewport=0bp 0bp 1282bp 382bp,width=1\columnwidth]{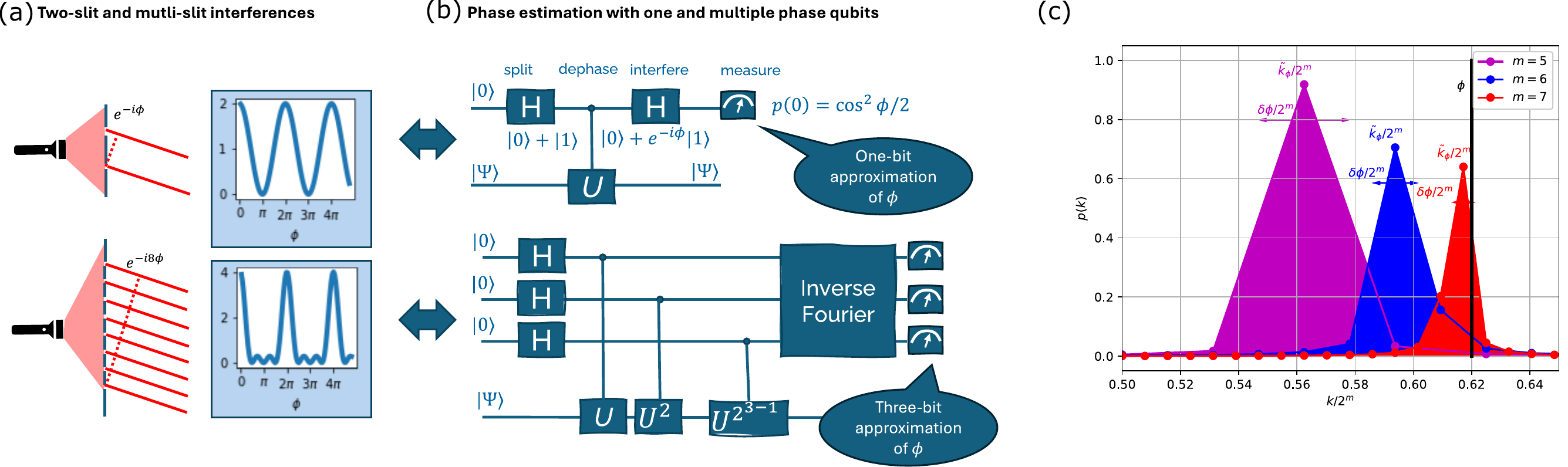}
\par\end{centering}
\caption{Phase estimation. Analogy between interferometry in optics (a) and
quantum phase estimation circuits (b), with $m=1$ ancilla qubit (two
slits, top) or $m=3$ ancilla qubits (8 slits, bottom). Adapted from
\cite{Ayral2025a}. (c) Histogram of the bitstring probabilities $p(j)=\left|\tilde{a}_{j}\right|$
for $\phi=E_{\alpha}t=0.62$ and a variable number $m$ of ancilla
qubits. \label{fig:quantum_phase_estimation}}
\end{figure}

\paragraph{Circuit implementation}

We now turn to the implementation of QPE\index{quantum phase estimation}.
QPE is essentially a multi-slit Mach-Zehnder interferometry experiment
\cite{Cleve1998}. It can be thought of as a Young slit experiment,
with a number of slits exponential in the number $m$ of ancilla bits.
Since interference experiments yield an accuracy scaling inversely
with the number of slits, this will give a $1/2^{m}$ accuracy (Fig.
\ref{fig:quantum_phase_estimation} (a)). Note that the $m=1$ circuit
reduces to the Hadamard test circuit (or two Young slits). .

The construction of the phase estimation circuit is illustrated in
Fig. \ref{fig:quantum_phase_estimation} (b). It consists in two groups
(registers) of qubits: the lower one, called the state register, for
preparing a state $|\Psi\rangle$ close to an eigenstate state $|\Psi_{\alpha}\rangle$,
and the upper one for $m$ ancilla qubits ($m=3$ on Fig. \ref{fig:quantum_phase_estimation}(b),
lower circuit), that will be used to read a $m$-bit estimate of $E_{\alpha}$. 

If $|\Psi\rangle=|\Psi_{\alpha}\rangle$, the state after the wall
of Hadamard gates and the controlled-$U^{2^{k-1}}$ ($k=1\dots m$)
gates is
\[
|\Psi_{\mathrm{tot}}\rangle=\bigotimes_{k=1}^{m}\left(\frac{|0\rangle+e^{iE_{\alpha}2^{k-1}t}|1\rangle}{\sqrt{2}}\right)\otimes|\Psi_{\alpha}\rangle=\frac{1}{2^{m/2}}\sum_{b_{1},\dots b_{m}}e^{iE_{\alpha}t\sum_{k=1}^{m}b_{k}2^{k-1}}|b\rangle\otimes|\Psi_{\alpha}\rangle.
\]
Defining $b=\sum_{k=1}^{m}b_{k}2^{k-1}$ (the integer representation
of the binary string $(b_{1},\dots,b_{m})$), we see that the state
before the inverse Fourier transform is $|\Psi_{\mathrm{tot}}\rangle=\sum_{b}a_{b}(E_{\alpha})|b\rangle\otimes|\Psi_{\alpha}\rangle$,
with $a_{b}(E_{\alpha})=\frac{1}{2^{m/2}}e^{iE_{\alpha}tb}$. We further
note that $a_{b+2\pi/(E_{\alpha}t)}=a_{b}$, if one assumes for the
sake of argument that $2\pi/(E_{\alpha}t)$ is an integer. That is,
the first part of the QPE circuit creates a state on the ancilla register
whose amplitudes is a periodic function of period $2\pi/(E_{\alpha}t)$.
The role of the Fourier transform is to extract this period by producing
the associated spectrum, which we expect to be peaked at a frequency
$\sim E_{\alpha}t$, from which we will extract the sought-after $E_{\alpha}$.

Let us check this intuition by doing the math: the inverse quantum
Fourier transform performs the (inverse) discrete Fourier transform
(DFT) of the amplitudes $a_{b}$: it yields ket $\sum_{j}\tilde{a}_{j}(E_{\alpha})|j\rangle$
on the ancilla register, with
\begin{align*}
\tilde{a}_{j}(E_{\alpha}) & =\mathrm{DFT}^{-1}\left[a_{b}(E_{\alpha})\right]=\frac{1}{\sqrt{2^{m}}}\sum_{b}e^{i2\pi bj/2^{m}}a_{b}(E_{\alpha})\\
 & =\frac{1}{2^{m}}\sum_{b=0}^{2^{m}-1}e^{i(2\pi j/2^{m}+E_{\alpha}t)b}=\frac{1}{2^{m}}\frac{\mathrm{sin}\left(\left(j+E_{\alpha}t\frac{2^{m}}{2\pi}\right)\pi\right)}{\mathrm{sin}\left(\left(j+E_{\alpha}t\frac{2^{m}}{2\pi}\right)2^{m}\pi\right)}.
\end{align*}
Defining $\phi=E_{\alpha}t$, this function is peaked at $j=k_{\phi}$
with $k_{\phi}=\left\lfloor 2^{m}\phi\right\rfloor $, as shown in
Fig. \ref{fig:quantum_phase_estimation} (c). We can thus take as
an estimate of the actual phase $\phi$ (and thus $E_{\alpha}$):
\begin{equation}
\hat{\phi}=\frac{k_{\phi}}{2^{m}}\label{eq:phase_decomp}
\end{equation}
with an error $\hat{\phi}-\phi=\delta\phi/2^{m}$: we see that we
are making an error $\delta\phi/2^{m}$ on the actual phase $\phi$:
QPE does yield an estimate of the phase (and hence $E_{\alpha}$)
to an $m$-bit precision, as advertised.

\paragraph{Resource estimation}

To conclude this subsection, let us evaluate the depth (number of
gates) of the QPE circuit, neglecting the state preparation: the dominant
contribution is given by the controlled-$U^{2^{k}}$ operations ($k=0$
to $m-1$). Each $U^{2^{k}}$ operation (neglecting the controls,
which will result in an additional overhead) will at least require
a depth of $2^{k}$ as per the non-fast-forwarding theorem ($U^{2^{k}}=e^{-iH(2^{k}t)}$
so the evolution time is $\propto2^{k}$), adding up to $O(2^{m})$
gates. The quantum Fourier transform requires $O(m^{2})$ gates, which
is negligible compared to the controlled evolutions. Here, we did
not specify the implementation of $U$ itself, but we could use Trotterization,
which will introduce a dependence of the gate count on the number
of terms $M$ and support of each term $s$ as per Eq. (\ref{eq:Ng_Trotter}),
and achieve a $t^{2}$ depth (at first order). Qubitization will achieve
a linear dependence in $t$, with additional dependence on the coefficients
of the sum-of-unitaries decomposition.

Let us emphasize that the $O(2^{m})$ depth is not to be considered
an exponential cost: it helps achieve an error $\epsilon=O(1/2^{m})$
on the estimate, so that the run time of QPE scales as $O(1/\epsilon)$,
a scaling known as ``Heisenberg scaling''. This scaling can be shown
to be optimal. It is quadratically better than the $1/\epsilon^{2}$
scaling of the Hadamard test method above, at the cost of a longer
circuit.

Let us finally point out that in both methods, the circuit depth (excluding
state preparation) scales polynomially with the system size, in contrast
to the classical methods that all come with some sort of exponential
scaling. As for the state preparation itself, it may require very
long circuits as discussed in subsection \ref{subsec:Ground-state-preparation:adiab}.

Alternative methods exist. \cite{Tong2020} uses the resolvent form
of the Green's function, Eq. (\ref{eq:resolvent_g_greater}). Using
this form, we can transform the problem of computing the Green's function
to the solution of a linear system of equations $Ax=b$ with $b=c_{\sigma}^{\dagger}|\Psi_{0}\rangle$
and $A=E_{0}-H+\omega+i\eta$ (typical algorithms for this task use
QPE as a subroutine). The overlap with solution $x=\frac{1}{E_{0}-H+\omega+i\eta}c_{\sigma}^{\dagger}|\Psi_{0}\rangle$
can then be computed with variations of the Hadamard test. Alternatively,
\cite{Tong2020} advocates the use of other advanced methods to compute
the matrix inverse appearing in the resolvent formula. 

\subsection{The exponential wall of decoherence: noisy quantum states and gates\label{subsec:The-exponential-wall}}

The algorithms presented above in principle allow for a tractable
(polynomial) computation of the (greater) Green's function in real
time or frequency, provided the ground or Gibbs state can be prepared...
and provided we are working with a perfect, that is, decoherence-free,
quantum computer. Here, we discuss the impact of decoherence.

\paragraph{Decoherence: density matrix and quantum channels}

In subsection \ref{subsec:Quantum-computing-in}, we described perfect
quantum computers. The physical implementations of quantum computers
that have become available in the recent years differ from this ideal
model. Their main constraint is decoherence, a phenomenon that stems
from the entanglement of the quantum computer with the outside world,
often called the environment. Denoting by $|\psi_{i}\rangle$ a basis
of the quantum computer's Hilbert space and $|\chi_{j}\rangle$ a
basis of said environment, the total wavefunction reads $|\Psi_{\mathrm{tot}}\rangle=\sum_{ij}c_{ij}|\psi_{i}\rangle\otimes|\chi_{j}\rangle$.
It is however impractical to work with such an object given the size
of the environment. All average observables on the quantum computer,
of the form $\langle O\rangle=\langle\Psi_{\mathrm{tot}}|\hat{O}\otimes I|\Psi_{\mathrm{tot}}\rangle$,
can be computed from the so-called density matrix $\rho=\mathrm{Tr}_{\mathrm{env}}\left(|\Psi_{\mathrm{tot}}\rangle\langle\Psi_{\mathrm{tot}}|\right)$
of the quantum computer (here $\mathrm{Tr}_{\mathrm{env}}$ denotes
the partial trace $\sum_{j}\langle\chi_{j}|\cdots|\chi_{j}\rangle$)
through the formula
\begin{equation}
\langle O\rangle=\mathrm{Tr}(\rho\hat{O}),\label{eq:avg_O_dens_mat}
\end{equation}
where the trace here is to be understood as the trace over the quantum
computer's Hilbert space. Therefore, for all practical purposes, quantum
states in the presence of decoherence will no longer be described
by a wavefunction $|\Psi\rangle$ but by a density matrix $\rho$,
a unit trace and positive definite matrix (as can be checked from
its definition). Of course, the density matrix also captures so-called
pure states (quantum states described by a single wavefunction $|\Psi\rangle$):
in a pure state has density matrix $\rho=|\Psi\rangle\langle\Psi|$.
Similarly, the time evolution of quantum states (which we conveniently
described with unitary gates) will no longer be described by a unitary
operator $|\Psi_{\mathrm{f}}\rangle=U|\Psi_{\mathrm{i}}\rangle$.
Instead we will consider linear, completely positive and trace-preserving
(CPTP) mappings $\mathcal{E}$ that act on the density matrix as follows:
\begin{equation}
\rho_{\mathrm{f}}=\mathcal{E}(\rho_{\mathrm{i}})=\sum_{k=1}^{K}E_{k}\rho_{\mathrm{i}}E_{k}^{\dagger},\label{eq:Kraus_channel}
\end{equation}
with $\sum_{k}E_{k}^{\dagger}E_{k}=I$. A noisy gate (also called
a quantum channel or CPTP map) is thus completely characterized by
$K$ operators $\left\{ E_{k}\right\} $ known as Kraus operators\index{decoherence}.
The trace preservation property ensures that the normalization of
the state ($\mathrm{tr}\rho=1$) is preserved, while complete positivity
means that applying the channel $\mathcal{E}$ on any subsystem (i.e
applying $\mathcal{E}\otimes\mathcal{I}$ with $\mathcal{I}$ the
identity channel) keeps the total density matrix positive definite
(applying a gate on, say, a single qubit should keep the many-qubit
quantum register physical). Note that when the Kraus rank $K$ is
one, we revert to the unitary case as $K_{1}^{\dagger}K_{1}=I$ and
$\rho_{\mathrm{f}}=K_{1}|\Psi_{\mathrm{i}}\rangle\langle\Psi_{\mathrm{i}}|K_{1}^{\dagger}=|\Psi_{\mathrm{f}}\rangle\langle\Psi_{\mathrm{f}}|$.

\paragraph{Standard error models}

Decoherence in quantum processors is usually described with a few
error metrics. An oft-used metric is the $T_{1}$ relaxation time
and the $T_{2}$ dephasing time, which are obtained through Rabi and
Ramsey experiments, respectively. From $T_{1}$, one can extract a
typical relaxation probability $p_{\mathrm{AD}}(T_{1})$ that enters
the quantum channel $\mathcal{E}_{\mathrm{AD}}$, dubbed amplitude
damping, that corresponds to this noise process. It is characterized
by the Kraus operators
\begin{equation}
K_{1}=\left[\begin{array}{cc}
1 & 0\\
0 & \sqrt{1-p_{\mathrm{AD}}}
\end{array}\right],\;\;K_{2}=\sqrt{p_{\mathrm{AD}}}\sigma_{-}.\label{eq:amplitude_damping_kraus}
\end{equation}
This means that with probability $p_{\mathrm{AD}}$, the state relaxes
from $|1\rangle$ to $|0\rangle$.

Similarly, from $T_{1}$ and $T_{2}$ one can compute a so-called
pure-dephasing time $T_{\varphi}=\left(1/T_{2}-1/(2T_{1})\right)^{-1}$,
which in turns determines a pure-dephasing probability $p_{\mathrm{PD}}(T_{\varphi})$
that enters the pure-dephasing quantum channel $\mathcal{E}_{\mathrm{PD}}$.
It is characterized by the Kraus operators
\begin{align}
K_{1} & =\sqrt{1-p_{\mathrm{PD}}}I,\;\;K_{2}=\sqrt{p_{\mathrm{PD}}}\sigma_{z}.\label{eq:pure_dephasing_kraus}
\end{align}
This means that with probability $p_{\mathrm{PD}}$, a superposed
state $|0\rangle+|1\rangle$ is dephased to $|0\rangle-|1\rangle$.
A similar channel, called the bit-flip channel (with $\sigma_{x}$
to replace $\sigma_{z}$ in $K_{2}$), corresponds to bit-flip errors.

Finally, let us introduce a widely used noise channel called the depolarizing
channel that, contrary to the pure-dephasing and bit-flip channels,
does not favor errors of one type (e.g $z$ or $x$) over others.
It is often defined (in the $N$-qubit case) as
\begin{equation}
\mathcal{E}_{\mathrm{D}}(\rho_{\mathrm{f}})=\left(1-p_{\mathrm{D}}\right)\rho_{\mathrm{f}}+p_{\mathrm{D}}I/2^{N}.\label{eq:def_depol_channel}
\end{equation}
Namely, with probability $p_{\mathrm{D}}$ the state is changed into
the so-called completely mixed state $I/2^{N}$, where all quantum
information is erased. (One can easily show that this channel has
four Kraus operators proportional to $I$, $\sigma_{x}$, $\sigma_{y}$
and $\sigma_{z}$, respectively, with the same coefficient in front
of the Pauli matrices, hence the claim that it does not favor one
error type over the others). The depolarizing channel is often used
for lack of more precise information as to the type of errors. For
instance, when one knows only the error rate of a certain unitary
gate $U^{(\mathrm{g})}$, one often models it as the composition$\mathcal{E}^{(g)}=\mathcal{E}_{\mathrm{D}}\circ\mathcal{E}_{\mathrm{perfect}}^{(g)}$,
with $\mathcal{E}_{\mathrm{perfect}}^{(g)}(\rho)=U^{(g)}\rho U^{(g)\dagger}$.

Let us end this section by examining the overall effect of noise on
a quantum circuit with $N_{\mathcal{\mathrm{g}}}$ gates that we assume
to be plagued by the same depolarizing noise $\mathcal{E}_{\mathrm{D}}$.
The final state can be written as
\begin{equation}
\rho_{\mathrm{f}}=\mathcal{E}^{(N_{\mathrm{g}})}\circ\mathcal{E}^{(N_{\mathrm{g}}-1)}\circ\cdots\circ\mathcal{E}^{(1)}(\rho_{\mathrm{i}}).\label{eq:final_noisy_state}
\end{equation}
Using the simple compositional noise model we just used, we easily
get the expression
\[
\rho_{\mathrm{f}}=\left(1-p_{\mathrm{D}}\right)^{N_{\mathrm{g}}}U|\Psi_{\mathrm{i}}\rangle\langle\Psi_{\mathrm{i}}|U^{\dagger}+\left(1-\left(1-p_{\mathrm{D}}\right)^{N_{\mathrm{g}}}\right)I/2^{N},
\]
with $U=U^{(N_{\mathrm{g}})}\cdot U^{(N_{\mathrm{g}}-1)}\cdots U^{(1)}$
the perfect circuit. We can readily evaluate the so-called fidelity
of the noisy final state with the perfect final state $|\Psi_{\mathrm{f}}\rangle=U|\Psi_{\mathrm{i}}\rangle$:
\begin{align}
\mathcal{F}(\rho_{\mathrm{f}},|\Psi_{\mathrm{f}}\rangle) & =\langle\Psi_{\mathrm{f}}|\rho_{\mathrm{f}}|\Psi_{\mathrm{f}}\rangle=\left(1-p_{\mathrm{D}}\right)^{N_{\mathrm{g}}}+\left(1-\left(1-p_{\mathrm{D}}\right)^{N_{\mathrm{g}}}\right)1/2^{N}\nonumber \\
 & \approx\left(1-p_{\mathrm{D}}\right)^{N_{\mathrm{g}}}\approx e^{-p_{\mathrm{D}}N_{\mathrm{g}}}.\label{eq:exponential_decay_fid}
\end{align}
The fidelity of the final state of a quantum circuit decays exponentially
with the number of gates $N_{\mathrm{g}}$ in this circuit. Conversely,
this means that for a given error rate $p_{\mathrm{D}}$ only circuits
of size $N_{\mathrm{g}}\lesssim1/p_{\mathrm{D}}$ are viable.

Today's error rates are between 1\% and 0.01\% for the limiting operations
(two-qubit gates), meaning only circuits that contain between 100
and 10,000 of these gates are viable. This, in practice, places very
severe constraints on the quantum circuits that can actually be run
on current and near-term processors. For instance, this rules out
quantum phase estimation circuits we discussed in subsection \ref{subsec:Frequency-domain:-quantum}.
Even simple Trotterization is limited to very short times and hence
poor energy resolutions. In fact, it already severely limits the preparation
of the ground state needed for the Green's function. In particular,
the use of adiabatic state preparation methods presented in subsection
\ref{subsec:Ground-state-preparation:adiab}, which require long circuits
due to the long annealing times, is prohibited.

To address this issue, new algorithms have been developed in the past
decade to reduce the number of requisite gates. This is the topic
of the next section.

\subsection{Near-term algorithms. Variational quantum algorithms for ground states
and for time evolution.\label{subsec:Near-term-algorithms.-Variationa}}

Current and near-term quantum processors come with drastic limitations
in terms of the number of gates that can be executed before decoherence
sets in. Starting in the mid-2010s, algorithms have been designed
for these ``noisy, intermediate-scale quantum'' (NISQ \cite{Preskill2018})
devices. 

\subsubsection{Ground state preparation: the variational quantum eigensolver}

\begin{figure}
\noindent \begin{centering}
\includegraphics[width=1\columnwidth]{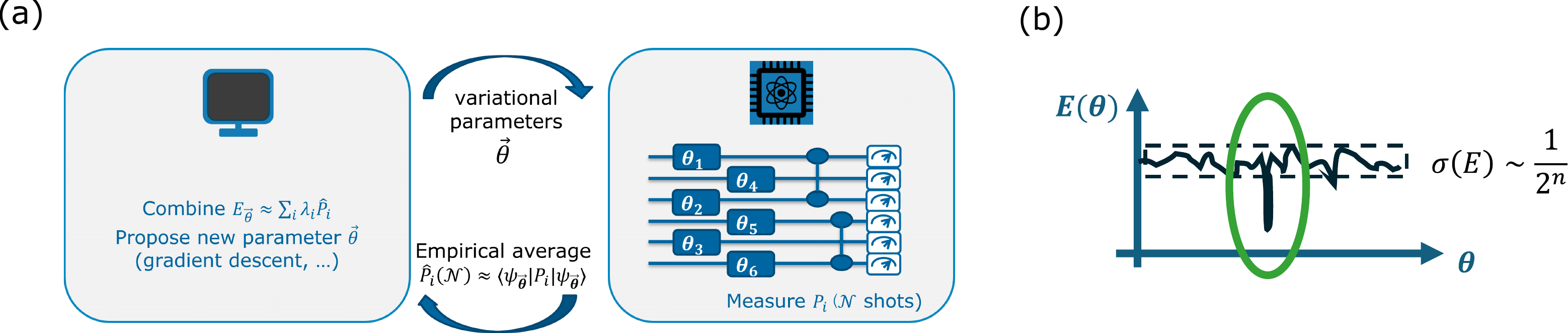}
\par\end{centering}
\caption{(a) Sketch of the variational quantum eigensolver (VQE) algorithm.
(b) Illustration of the barren plateau problem: the variance vanishes
exponentially on average.\label{fig:Sketch-of-the-VQE}}
\end{figure}

The most prominent example is the variational quantum eigensolver
(VQE\index{variationan quantum eigensolver}, \cite{Peruzzo2014},
see \cite{Tilly2021a} for a review), which is used to prepare ground
states of Hamiltonians and measure their energy. Like DMRG above (paragraph
\ref{par:Matrix-product-states:}), VQE is a variational method that
sets out to minimize a variational energy $E_{\boldsymbol{\theta}}=\langle\Psi(\boldsymbol{\theta})|H|\Psi(\boldsymbol{\theta})\rangle$,
with $|\Psi(\boldsymbol{\theta})\rangle$ a family of variational
states (also called ``ansatz''). The specificity of VQE is that
the variational states are generated on a quantum processor by a unitary
operation $U(\boldsymbol{\theta})$:
\[
|\Psi(\boldsymbol{\theta})\rangle=U(\boldsymbol{\theta})|\Phi_{\mathrm{init}}\rangle,
\]
where $|\Phi_{\mathrm{init}}\rangle$ is usually state $|0\rangle\otimes|0\rangle\cdots\otimes|0\rangle,$
and $U(\boldsymbol{\theta})$ represents a quantum circuit with parameters
$\boldsymbol{\theta}$ (that may be, for instance, the angles of rotation
gates $R_{\alpha}(\theta)$). Then, the energy $E_{\boldsymbol{\theta}}$
of this state is measured on the quantum computer. As explained in
subsection \ref{subsec:Quantum-measurements-and}, this measurement
can be done by achieving a Pauli decomposition of $H$ (see Eq. (\ref{eq:Pauli_decomp_H}))
and then summing up each contribution:
\begin{equation}
E_{\boldsymbol{\theta}}=\sum_{i=1}^{M}\lambda_{i}\langle\Psi(\boldsymbol{\theta})|P_{i}|\Psi(\boldsymbol{\theta})\rangle.\label{eq:variational_energy}
\end{equation}
Each expectation value $\langle\Psi(\boldsymbol{\theta})|P_{i}|\Psi(\boldsymbol{\theta})\rangle$
is then estimated by an empirical average over $N_{\mathrm{s}}$ shots.
Based on the computed $E_{\boldsymbol{\theta}}$, a classical optimizer
updates the parameters $\boldsymbol{\theta}$. This is summarized
in Fig. \ref{fig:Sketch-of-the-VQE} (a). VQE comes with three main
challenges: decoherence, the measurement of $E_{\boldsymbol{\theta}}$,
and the variational optimization of the parameters.

\paragraph{Decoherence and the design of ans�tze}

Decoherence, as we just saw, limits the size of the circuit $U(\boldsymbol{\theta})$
that can be run in VQE. The main advantage of VQE is its flexibility
in terms of ansatz: given a fixed gate budget, one is free to pick
the ansatz that best fits the problem at hand. Ansatz construction
strategies fall into two main categories: hardware-efficient ans�tze
\cite{Kandala2017} and physics-inspired ones. The former aims at
achieving the most expressivity given the gates available on the quantum
processor. Typically, hardware-efficient ans�tze feature a repetition
of the following pattern: a sequence of one qubit rotations applied
to each qubit, followed by the hardware's native two-qubit gate (there
is usually only one) applied to all qubits pairs compatible with the
connectivity of the qubits. The rotations introduce variational parameters,
while the two-qubit gates generate entanglement. In fact, this construction
guarantees a fast growth of entanglement---an important feature given
the limited coherence time. However, it does not take into account
the specifics of the Hamiltonian at hand. This is what the latter
approach does.

For instance, the Hamiltonian variational ansatz (HVA, \cite{Wecker2015})
draws inspiration from the adiabatic quantum annealing algorithm (subsection
\ref{subsec:Ground-state-preparation:adiab}): based on the intuition
that one should start with the ground state $|\Phi_{0}\rangle$ of
a simple Hamiltonian $H_{0}$ and deform the latter to the Hamiltonian
at hand $H$, it uses a Trotterized form of the time evolution
\[
|\Psi(t_{\mathrm{annealing}})\rangle=U(t_{\mathrm{annealing}})|\Phi_{0}\rangle=\prod_{k=1}^{M}e^{-i\left(1-t_{k}/t_{\mathrm{annealing}}\right)H_{0}t_{k}}e^{-it_{k}/t_{\mathrm{annealing}}Ht_{k}}|\Phi_{0}\rangle
\]
(for $t_{k}=kt_{\mathrm{annealing}}/M$) as an inspiration to the
following variational form:
\begin{equation}
|\Psi(\boldsymbol{\theta})\rangle=\prod_{k=1}^{M}e^{-i\theta_{2k}H_{0}}e^{-i\theta_{2k+1}H}|\Phi_{0}\rangle\label{eq:U_HVA}
\end{equation}
with $2M$ parameters $\boldsymbol{\theta}$. (Then the operators
$e^{-i\theta_{2k}H_{0}}$ and $e^{-i\theta_{2k+1}H}$ can be translated
to quantum circuits using the Trotterization tools we introduced in
subsection \ref{subsec:Time-evolution:Trotter}). In the limit of
large $M$, the assignment $\theta_{2k}=(1-t_{k}/t_{\mathrm{annealing}})t_{k}$
and $\theta_{2k+1}=t_{k}^{2}/t_{\mathrm{annealing}}$ of the parameters
guarantees a convergence to the ground state. However, the premise
of HVA is that the variational optimization of $\boldsymbol{\theta}$
can lead to a good enough approximation of the ground state in a much
shorter time (that is, compatible with the coherence time)... providing
a sort of shortcut to adiabaticity.

Typically, in the context of the AIM (Eq. (\ref{eq:AIM_hamiltonian})),
a natural choice of $H_{0}$ is $H_{0}=Un_{\uparrow}n_{\downarrow}-\mu\sum_{\sigma}c_{\sigma}^{\dagger}c_{\sigma}+\sum_{k\sigma}\epsilon_{k}a_{k\sigma}^{\dagger}a_{k\sigma}$
because its ground state, after a Jordan-Wigner encoding, is a computational
basis state that can be prepared with $\sigma_{x}$ gates only. Then,
one can replace $H$ with $H'=\sum_{k\sigma}V_{k}\left(a_{k\sigma}^{\dagger}c_{\sigma}+\mathrm{h.c}\right)$.

Other approaches are targetted more specifically at fermionic states:
for instance, the low-depth circuit ansatz (LDCA, \cite{Dallaire-Demers2018})
proposes to intercalate so-called Gaussian gates (which generate only
Gaussian states, namely states generated by Hamiltonians quadratic
in the $c$ and $c^{\dagger}$ operators) with non-Gaussian gates
(corresponding to quartic terms or beyond).

Once the ansatz is picked, other optimizations to reduce the circuit
depth and thus counter decoherence is the choice of the orbital basis
in which the states are expressed. For instance, while a Fock state
$\prod_{k=1}^{N}\left(c_{k}^{\dagger}\right)^{n_{k}}|0,\dots,0\rangle$
expressed in the Fock basis can be generated, upon Jordan-Wigner encoding,
simply by applying $\sigma_{x}$ gates to the qubits with occupations
$n_{k}=1$, if the same state is expressed in any other basis, longer
circuits (possibly including entangling gates) will be necessary.
For a generic state $|\Psi\rangle$, the choice of the natural-orbital
basis (that diagonalizes the one-particle reduced density matrix $D_{ij}=\langle\Psi|c_{i}^{\dagger}c_{j}|\Psi\rangle$)
is supposed to yield the most compact representation of the state
and should therefore require shorter quantum circuits (\cite{Besserve2021,Besserve2024,Besserve2024a}).

Finally, one can also build the ansatz iteratively instead of starting
from a fixed ansatz: this is what the ADAPT-VQE method (\cite{Grimsley2019,Tang2019})
does. At each step, it picks from a pool of gates the gate that maximizes
the gradient of the energy with respect to the gate's parameter, therefore
ensuring the fastest convergence to the minimum.

A major mostly uncharted territory is the design of ans�tze specific
to the nature of the AIM (beyond what the HVA does). In particular,
the fact that the bath is non-interacting, which allows for drastic
simplifications in action-based impurity solvers (subsection \ref{subsec:Action-based-solvers}),
has not been exploited so far.

\paragraph{The measurement problem}

VQE resorts to a classical summation of individual samples to compute
the variational energy $E_{\boldsymbol{\theta}}$ (Eq. (\ref{eq:variational_energy})).
This means that the statistical error will scale as $\epsilon=O(1/\sqrt{N_{\mathrm{s}}})$,
with $N_{\mathrm{s}}$ the number of samples (shots). As explained
before, this scaling of $1/\epsilon^{2}$ of the run time is much
less favorable than the $1/\epsilon$ scaling of quantum phase estimation.

The prefactor of this scaling can be mitigated by many methods. One
can for instance group terms that commute with one another: they can
be measured without regenerating the state. One can also allocate
the total shot budget based on the importance of each term in the
Pauli decomposition: larger coefficients $|\lambda_{i}|$ in Eq. (\ref{eq:variational_energy})
should receive more shots. The optimal shot allocation can be worked
out (\cite{Rubin2018}, and see \cite{Arrasmith2020} for an adaptive
shot allocation strategy), leading to a standard error on the mean
that satisfies the upper bound:
\begin{equation}
\Delta E_{\boldsymbol{\theta}}\leq\frac{\sum_{i=1}^{M}\left|\lambda_{i}\right|}{\sqrt{N_{\mathrm{s}}}}.\label{eq:shot_noise_bound}
\end{equation}
The numerator of the upper bound is referred to as the one-norm $H$.
Proposals have been made to minimize it by orbital rotations \cite{Koridon2021}.

Another method called classical shadows \cite{Huang2020} is based
on the realization that measuring $E_{\boldsymbol{\theta}}$ by measuring
each $P_{i}$ (or group thereof) individually on the final state $|\Psi(\boldsymbol{\theta})\rangle$
(or $\rho_{\boldsymbol{\theta}}$ in the general case where it is
afflicted by noise) is a poor reuse of the information in $\rho_{\boldsymbol{\theta}}$.
Classical shadows thus first produce an estimate (called classical
shadow) $\hat{\rho}_{\boldsymbol{\theta}}$ of $\rho_{\boldsymbol{\theta}}$
by appending gates drawn randomly for smartly chosen groups of unitaries
to the circuit and then measuring along the $\sigma_{z}$ axis...
and then return estimates $\langle\hat{P}_{i}\rangle=\mathrm{Tr}(\hat{\rho}_{\boldsymbol{\theta}}P_{i})$
to compute the energy $E_{\boldsymbol{\theta}}$. By so doing, the
statistical error is still behaving as $1/\sqrt{N_{\mathrm{s}}}$(as
per the central limit theorem), but the numerator can be mitigated
(achieving, in some cases, a $\log(M)$ scaling, with $M$ the number
of Pauli terms). Adaptations to fermionic settings have appeared recently,
albeit not in a VQE context \cite{Wan2023}.

\paragraph{The optimization problem}

Supposing the ansatz is short enough to beat decoherence, and the
energy $E_{\mathrm{\boldsymbol{\theta}}}$ can be measured to enough
statistical accuracy, the VQE practitioner is still facing the problem
of finding the minimum of $E_{\boldsymbol{\theta}}.$ The difficulty
of this optimization task has been a major focus of recent years.

On the practical side, methods have been developed to measure the
gradients required in gradient-based minimization methods, like the
gradient descent method, which updates parameters through
\begin{equation}
\boldsymbol{\theta}_{i+1}=\boldsymbol{\theta}_{i}-\alpha\boldsymbol{\nabla}_{\theta}E(\boldsymbol{\theta}_{i}),\label{eq:gradient_descent}
\end{equation}
with $\alpha$ a learning rate and $\left[\boldsymbol{\nabla}_{\theta}E(\boldsymbol{\theta})\right]_{k}=\frac{\partial E(\boldsymbol{\theta})}{\partial\theta_{k}}=\frac{\partial}{\partial\theta_{k}}\langle\Psi_{\theta}|H|\Psi_{\theta}\rangle.$
It turns out that if the parameteric gates used in the ansatz are
of the form $U_{k}(\theta_{k})=e^{-i\theta_{k}/2P_{k}}$, with $P_{k}^{2}=I$,
then the following identity (called ``parameter shift rule''\index{parameter shift rule})
can be easily derived:

\begin{equation}
\frac{\partial E(\boldsymbol{\theta})}{\partial\theta_{k}}=\frac{1}{2}\left(E(\theta_{k}+\pi/2)-E(\theta_{k}-\pi/2)\right),\label{eq:parameter_shift_rule}
\end{equation}
where $E(\theta_{k}+\alpha)$ is shorthand for $E(\theta_{1},\dots,\theta_{k}+\alpha,\dots,\theta_{L})$.
Estimating gradients is therefore relatively straightforward. It however
requires a number of evaluations proportional to the number $L$ of
parameters, a major overhead compared to the $O(1)$ cost of computing
gradients in classical deep neural networks using the backpropagation
algorithm \cite{Atilim2017}. Other machine-learning-inspired methods
like natural gradients can also be implemented \cite{Stokes2020},
albeit at a much larger cost since the required quantum geometric
tensor necessitates $O(L^{2})$ energy evaluations.

On the more formal side, the potential landscape $E_{\boldsymbol{\theta}}$
(and more specifically its variance) has been studied for some classes
of variational circuits $U_{\boldsymbol{\theta}}$ and observables
$H$. The empirically observed vanishing of gradients in large regions
of the variational space, dubbed ``barren plateau problem'' \cite{McClean2018}
is now understood as a curse of dimensionality issue \cite{Ragone2023,Larocca2024}:
the larger the dimension of the effective space explored by the variational
search, the more concentrated $E_{\boldsymbol{\theta}}$ around its
mean and therefore the smaller the gradients (in average, as illustrated
in Fig. \ref{fig:Sketch-of-the-VQE} (b)). This issue is particularly
pressing for the hardware-efficient parametrized circuits we introduced
above: if they are deep enough, the ensemble of unitaries $U_{\boldsymbol{\theta}}$
is very close to the group of Haar random matrices and the variance
$\langle\left(E_{\boldsymbol{\theta}}-\langle E_{\boldsymbol{\theta}}\rangle\right)^{2}\rangle$
over the parameter landscape vanishes exponentially with the number
$N$ of qubits. In other words, the more expressive, the less trainable
parametrized circuits become. On the other hand, physics-inspired
ans�tze, which usually come with more structure, could suffer less
from this issue. Note also that the existence of barren plateaux is
an average phenomenon: on average, energies are (exponentially) close
to their mean. However, non-zero gradient portions of the parameter
space exist (see circled region in \ref{fig:Sketch-of-the-VQE} (b)).
The challenge lies in finding them. In particular, finding a good
starting point $\boldsymbol{\theta_{0}}$ for the optimization appears
crucial. There again, physics-informed choices (by e.g perturbation
theory or classical pretraining of the variational ansatz) could play
a crucial role. Adaptive techniques like the aforementioned ADAPT-VQE
methods, or other ways of solving the eigenvalue problem (see e.g
\cite{Plazanet2024}) could also mitigate this issue. 

\subsubsection{Time evolution: variational circuits for time evolution.}

The Trotter circuits described in a previous section are often too
long compared to the available number of gates given the coherence
time. Methods, like ``variational quantum simulation'' \cite{Endo2019},
have been designed to train variational circuits to perform a given
time evolution. They are a simple adaptation of the variational principle
used in VQE to the time-dependent case: given a variational state
$|\Psi(\boldsymbol{\theta}(t))\rangle=U(\boldsymbol{\theta}(t))|\Psi_{\mathrm{init}}\rangle$
($|\Psi_{\mathrm{init}}\rangle=c^{\dagger}|\Psi_{0}\rangle$ for the
greater Green's function), one can use the McLachlan variational principle
\[
\min\delta\left\Vert \left(\frac{\mathrm{d}}{\mathrm{dt}}+iH\right)|\Psi(\boldsymbol{\theta}(t))\rangle\right\Vert 
\]
to write a differential equation for the parameters $\boldsymbol{\theta}(t)$:
\[
\sum_{j}M_{ij}\dot{\theta}_{j}=V_{i}
\]
with $M_{i,j}=\mathrm{Re}\left(\frac{\partial\langle\Psi(\boldsymbol{\theta}(t))|}{\partial\theta_{i}}\frac{\partial|\Psi(\boldsymbol{\theta}(t))\rangle}{\partial\theta_{j}}\right)$
and $V_{i}=\mathrm{Im}\left(\langle\Psi(\boldsymbol{\theta}(t))|H\frac{\partial|\Psi(\boldsymbol{\theta}(t))\rangle}{\partial\theta_{i}}\right)$.
Both quantities can be evaluated with dedicated quantum circuits similar
to the Hadamard test \cite{Mitarai2019a,McArdle2018}. Interestingly,
the aforementioned natural gradient approach can be obtained by invoking
the McLachlan principle with imaginary times.

\subsubsection{Reducing the effect of imperfections: error mitigation.}

All the methods above aim at creating shorter circuits, but still
suffer from decoherence. A variety of techniques known as error mitigation
has been developed over the years to counter the exponential decay
of fidelity (Eq. (\ref{eq:exponential_decay_fid})). They generically
trade a systematic bias generated by decoherence for an increased
classical cost---that essentially helps recover the lost information.

A representative example of this tradeoff is probabilistic error cancellation
\cite{Endo2017,Temme2017}. To suppress the bias in energy in a noisy
VQE, caused by the deviation between the noisy energy $E_{\mathrm{noisy}}=\mathrm{Tr}(\rho_{\mathrm{f}}H)$,
with $\rho_{\mathrm{f}}$ given by Eq. (\ref{eq:final_noisy_state}),
and the perfect energy $E_{\mathrm{perfect}}=\langle\Psi_{\mathrm{f}}|H|\Psi_{\mathrm{f}}\rangle$,
one decomposes the ``perfect'' channels $\mathcal{E}_{\mathrm{perfect}}^{(k)}(\rho)$
as a linear combination of the actual (noisy) channels $\mathcal{E}^{(k)}(\rho)$
that are realized in the experiment:
\begin{equation}
\mathcal{E}_{\mathrm{perfect}}^{(k)}=\sum_{l}q_{l}^{(k)}\mathcal{E}_{l},\label{eq:lc_channels_pec}
\end{equation}
with the underlying assumption that these noisy channels form an independent
family, and the $\sum_{l}q_{l}^{(k)}=1$ to preserve the trace-preserving
character (but the $q_{l}^{(k)}$ may be negative). The perfect energy
can thus be decomposed as a very large sum, which is sampled through
Monte-Carlo:
\begin{align}
E_{\mathrm{perfect}} & =\sum_{l_{1}\dots l_{N_{\mathrm{g}}}}q_{l_{1}}^{(1)}\cdots q_{l_{N_{\mathrm{g}}}}^{(N_{\mathrm{g}})}\mathrm{Tr}\left(H\mathcal{E}_{l_{N_{\mathrm{g}}}}\circ\mathcal{E}_{l_{N_{\mathrm{g}}-1}}\circ\cdots\mathcal{E}_{l_{1}}(\rho_{\mathrm{i}})\right)\nonumber \\
 & =\Gamma\sum_{l_{1}\dots l_{N_{\mathrm{g}}}}p_{l_{1}}^{(1)}\cdots p_{l_{N_{\mathrm{g}}}}^{(N_{\mathrm{g}})}\mathrm{s}(l_{1},\dots l_{N_{\mathrm{g}}})\mathrm{Tr}\left(H\mathcal{E}_{l_{N_{\mathrm{g}}}}\circ\mathcal{E}_{l_{N_{\mathrm{g}}-1}}\circ\cdots\mathcal{E}_{l_{1}}(\rho_{\mathrm{i}})\right)\\
 & \approx\frac{\Gamma}{N_{s}}\sum_{i=1}^{N_{s}}\mathrm{s}(l_{1}^{(i)},\dots l_{N_{\mathrm{g}}}^{(i)})\mathrm{Tr}\left(H\mathcal{E}_{l_{N_{\mathrm{g}}}^{(i)}}\circ\mathcal{E}_{l_{N_{\mathrm{g}}-1}^{(i)}}\circ\cdots\mathcal{E}_{l_{1}^{(i)}}(\rho_{\mathrm{i}})\right).\label{eq:pec_formula}
\end{align}
Here, the probabilities $p_{l}^{(k)}=q_{l}^{(k)}/\sum_{l'}q_{l'}^{(k)}$
are introduced to deal with the possible negativity of the $q_{l}^{(k)}$'s,
$\Gamma=\prod_{k=1}^{N_{g}}\sum_{l'}\left|q_{l'}^{(k)}\right|$ and
$\mathrm{s}(l_{1},\dots l_{N_{\mathrm{g}}})=\prod_{k=1}^{N_{g}}\mathrm{sign}(q_{l_{k}}^{(k)})$.
The advantage of the summand of Eq. (\ref{eq:pec_formula}) is that
it can be estimated with the noisy computer at hand by running a circuit
with (noisy) gates $\mathcal{E}_{l_{1}^{(i)}},\dots,\mathcal{E}_{l_{N_{\mathrm{g}}}^{(i)}}$
and measuring $H$ at the end, resulting in an unbiased estimate of
$E_{\mathrm{perfect}}$! The price to pay is, however, twofold: one
needs to know precisely the noise models of the hardware to be able
to perform the decomposition of Eq. (\ref{eq:lc_channels_pec}), which
requires quite expensive process tomography experiments. More fundamentally,
the estimator comes with a statistical uncertainty that scales with
the $\Gamma$ factor, which can be rewritten as $\Gamma=\prod_{k=1}^{N_{g}}\left(1+2\eta_{k}\right)$,
with the so-called ``negativity'' $\eta_{k}=\sum_{l,q_{l}<0}q_{l}^{(k)}$.
To get an intuition how it scales, let us suppose that all gates have
the same negativity $\eta$. Then
\begin{equation}
\Gamma\approx e^{2\eta N_{\mathrm{g}}}.\label{eq:negativity_exponential}
\end{equation}
In other words, to get a fixed statistical error bar $\epsilon\propto\Gamma/\sqrt{N_{\mathrm{s}}}$,
one needs to scale the number $N_{\mathrm{s}}$ of samples as $\Gamma^{2}=e^{4\eta N_{\mathrm{g}}}$,
namely exponentially with the number of gates! This essentially tells
us that to fight against the exponential loss in fidelity (Eq. (\ref{eq:exponential_decay_fid})),
we need to pay an exponential price. In practice, this exponential
can be manageable provided the negativity is not too large. The better
the hardware, the closer $\eta$ to unity, the lower the sampling
overhead.

\section{Conclusion: state of the art, challenges and ways ahead}

\begin{table}
\begin{tabular}{|c|>{\centering}p{0.7cm}|>{\centering}p{0.7cm}|>{\centering}p{2.5cm}|>{\centering}p{3cm}|>{\centering}p{1cm}|>{\centering}p{1.5cm}|>{\centering}p{3cm}|}
\hline 
Ref. & $N_{\mathrm{c}}$ & $N_{\mathrm{b}}$ & State preparation  & Green's function method & Noisy emulation & Physical implem. & Remark\tabularnewline
\hline 
\hline 
\cite{Bauer2016} & any & any & Adiabatic state preparation & Hadamard test & No & No & Generic proposal.\tabularnewline
\hline 
\cite{Kreula2016} & 1 & 10 & N-A & Hadamard (Trotter) & Yes & Ions (th) & Nonequilibrium: kinetic ramp\tabularnewline
\hline 
\cite{Kreula2016a} & 1 & 1 & Exact & Hadamard with Trotter & Yes & SC (th) & Study of noise in bath.\tabularnewline
\hline 
\cite{Rungger2019} & 1 & 1 & VQE (HEA) & Lehmann (excited-VQE ) & No & SC + Ions (exp) & No self-consistency\tabularnewline
\hline 
\cite{Jaderberg2020} & 1 & 1 & Exact and VQE & Hadamard (Trotter) & Yes & No & Variational compression.\tabularnewline
\hline 
\cite{Keen2019} & 1 & 1 & VQE (adhoc) & Hadamard (Trotter) & No & SC (exp) & \tabularnewline
\hline 
\cite{Yao2020} & 1 & 1 & VQE (UCC) & N/A & No & SC (exp) & Periodic Anderson model. Parity mapping\tabularnewline
\hline 
\cite{Steckmann2021} & 1 & 1 & VQE (adhoc) & Hadamard (fast-forwarding) & No & SC (exp) & \tabularnewline
\hline 
\cite{Besserve2021} & 2 & 2 & VQE (adhoc) & N/A & Yes & No & RISB.\tabularnewline
\hline 
\cite{Jamet2021} & 1 & 3 & VQE (HEA) & Krylov variational algorithm & No & No & QSGW+DMFT La$_{2}$CuO$_{4}$\tabularnewline
\hline 
\cite{Jamet2022} & 1 & 7 & Quantum subspace expansion & Quantum subspace expansion & No & No & \tabularnewline
\hline 
\cite{Ehrlich2023} & 1 & 1 & VQE (symmetric HEA) & Lehmann (excited-VQE) & Yes & SC (exp) & \tabularnewline
\hline 
\cite{Jamet2023} & 3 & 17 & Matrix product states & Quantum subspace expansion & No & No & SrVO$_{3}$, chain geometry.\tabularnewline
\hline 
\cite{Baul2024} & 1 & 1 & VQE (\cite{Keen2019}) & Hadamard (Trotter) & No & No & \tabularnewline
\hline 
\cite{Selisko2024} & 1 & 6 & VQE (HEA, classical optim.) & Lehmann (qEOM) & No & SC (exp) & Ca$_{2}$CuO$_{2}$Cl$_{2}$, chain geometry\tabularnewline
\hline 
\cite{Jones2024} & 1 & 6 & VQE (sym.-preserving ansatz) & Krylov variational algorithm & No & No & \tabularnewline
\hline 
\end{tabular}

\caption{Summary of early implementations of impurity solvers with quantum
computers, sorted in chronological order. $N_{\mathrm{c}}$: number
of impurities. $N_{\mathrm{b}}$: (maximum) number of bath sites.
SC stands for superconducting qubits. All DMFT computations except
those on materials are done on the Bethe lattice. The encoding is
always Jordan-Wigner, unless otherwise stated.\label{tab:DMFT-on-QC}}
\end{table}

The tools we just introduced have been used in the past 10 years to
solve the impurity model of DMFT with the help of quantum processors.
Table \ref{tab:DMFT-on-QC} summarizes the various attempts. After
the original proposal by \cite{Kreula2016} and \cite{Bauer2016},
most works (\cite{Kreula2016a,Rungger2019,Jaderberg2020,Keen2019,Steckmann2021,Ehrlich2023,Baul2024})
focused on the simplest DMFT scheme, namely two-site DMFT \cite{Potthoff2001},
which consists in limiting the bath to only one bath site, requiring
only 4 qubits. Most works used VQE in combination with the Hadamard
test circuit we introduced before, with various ans�tze and methods
to optimize the time evolution needed in the Hadamard test. The Hadamard
test method is replaced by a Lanczos-type algorithm in \cite{Jamet2021,Jones2024}
that builds the Krylov states using a variational ansatz $\text{|\ensuremath{\chi_{n}\rangle=U(\boldsymbol{\theta}_{n})|\Phi_{\mathrm{init}}\rangle}}$.

Driven by the optimization issues of variational methods, quantum
subspace expansions are also proposed, in an impurity model context,
by \cite{Jamet2022}: the Krylov bases (whether for the ground state
or for the Green's function, see paragraph \ref{par:Exact-diagonalization:-Lanczos})
are constructed not in the full Hilbert space but in a reduced subspace
built with states obtained by Trotter iterations: $|\psi_{k}\rangle=U(\delta t)^{k}|\Phi_{\mathrm{init}}\rangle$,
with $U(\delta t)=e^{-iH_{\mathrm{AIM}}\delta t}$. The conjecture
of the method is that Trotter iterates will yield a good enough subspace
of the full Hilbert space to perform the Lanczos method in. The first
(ground state) Lanczos iteration can also be replaced by a purely
classical method, as advocated in \cite{Jamet2023}: there, DMRG is
used to compute the ground state in the matrix product state form.
The MPS is converted to a quantum circuit and then Trotter iterates
$|\psi_{k}\rangle$ are generated on top of this state.

The above quantum Krylov variants require the computation of various
matrix elements of the Trotter iterates $|\psi_{k}\rangle$, which
incurs the aforementioned $1/\epsilon^{2}$ overhead. \cite{Yu2025}
proposes a way to avoid this overhead by simply sampling bitstrings
$s_{i,k}$ from the $|\psi_{k}\rangle$'s and building the classical
representation of $H$ in the subspace corresponding to these bitstrings.
\cite{Yu2025} then classically finds the ground state energy by diagonalizing
the restriction of $H$ to this subspace for systems of up to 41 bath
sites (but does not compute the Green's function).

\cite{Kreula2016a} looks at a nonequilibrium setting, which requires
extending the formalism we introduced to Keldysh Green's functions.

Alternative embedding techniques---which can be regarded as low-energy
approximations of DMFT \cite{Ayral2017a}---have been used, motivated
by the fact that these techniques require an impurity problem with
fewer bath sites ($N_{b}=N_{c}$), and require not the full time-dependent
Green's function, but only the one-particle reduced density matrix
(essentially the $t=0$ Green's function). This includes rotationally-invariant
slave bosons (aka Gutzwiller \cite{Yao2020,Besserve2021}) and density-matrix
embedding theory \cite{Rubin2016,Ma2020,Cao2022}. Finally, simpler
slave-particle methods like $\mathbb{Z}_{2}$ slave spins were used
to reduce the Hubbard model to even simpler (namely spin-based, as
opposed to fermionic) effective models \cite{Michel2023a}.

The current state of affairs is that the small sizes considered in
the aforementioned publications are not yet large enough (except perhaps
for \cite{Jamet2023,Yu2025}) that ground state preparation issues
with VQE (like barren plateaux) become severe, so that whether a variational
preparation of low-energy states of the AIM with quantum methods can
succeed is still an open question. As for the time evolution needed
to compute Green's function, the decoherence rates of current processors
has so far prevented one from reaching numbers of bath sites competitive
with the best classical methods: larger bath sizes mean that larger
evolution times, and thus larger circuits, are needed, at least with
the Hamiltonian representation of the impurity model we worked with
so far.

The bath size of a Hamiltonian-based representation of the AIM of
DMFT will likely remain a major issue. Large bath sizes (many Dirac
delta peaks, as in (\ref{eq:fit_problem})) are needed to represent
the spectral features of the hybridization function, and these in
turn entail large numbers of qubits, and thus large circuits. From
a nonequilibrium perspective, one can also argue that reaching long
times require large baths to avoid finite-size effects. In principle,
the fact that the bath is noninteracting could be exploited to simplify
circuits or perhaps reduce the number of qubits, but this has not
been attempted yet, to the best of our knowledge---other than switching
from DMFT to simpler embedding techniques like RISB and DMET, which
can be seen as well-defined prescriptions for truncating the bath,
albeit with access only to one-particle reduced density matrices (as
opposed to frequency-dependent Green's functions).

An alternative, promising route is to use another representation of
the impurity model. For instance, open-system representations of the
impurity model have been proposed \cite{Arrigoni2012,Dorda2016,Schwarz2016}
that use a dissipative bath instead of a noninteracting closed bath:
each bath site can exchange electrons with an environment. The rate
of dissipation $\Lambda$ (as well as the usual other hybridization
parameters) can be fixed by matching the hybridization function not
with a sum of Dirac peaks (Eq. (\ref{eq:fit_problem})) but of Lorentzians
of width $\Lambda$:
\begin{equation}
\Delta^{\mathrm{R}}(\omega)=\sum_{k}\frac{V_{k}^{2}}{\omega+i\Lambda-\epsilon_{k}}.\label{eq:fit_function_noisy}
\end{equation}
On a classical processor, this allows to use fewer peaks to fit the
same hybridization function (with many possible sophistications of
the open-system representation \cite{Huang2025}), and then, using
a master equation-based solver, compute the corresponding Green's
function. Alternatively, one could use a noisy quantum processor to
perform the corresponding dissipative time evolution \cite{Bertrand2024}.
This way, the natural decoherence of the processor could, at least
partly, be used to mimic the physics of impurity electrons in solids.

This open-system representation may also solve a deeper issue a quantum
algorithms for impurity models, namely the preparation of a ground
(or low-energy) state. Working with dissipative models changes the
perspective to preparing the steady state of an open quantum system...
In such a context, more noise could help reach this steady state faster.

\section*{Acknowledgments}
We thank Fran\c cois Jamet for a careful reading of the manuscript.

\clearpage{}

\bibliographystyle{correl}

\end{document}